\documentclass[prb,twocolumn,superscriptaddress,showpacs,amsmath,amssymb,longgraphy]{revtex4-1}
\usepackage{graphicx, color}
\usepackage{dcolumn}
\usepackage{bm}
\usepackage{epstopdf}
\usepackage{physics}
\usepackage[usenames,dvipsnames]{xcolor}
\usepackage{subfigure}
\usepackage{pifont}
\usepackage{appendix}
\usepackage{subfig}
\usepackage{lipsum}

\usepackage{verbatim}

\begin{document}

\author{I. V. Tokatly}
\email{ilya.tokatly@ehu.es}
\affiliation{Nano-Bio Spectroscopy Group, Departamento de Polímeros y Materiales Avanzados, Universidad del País Vasco, 20018 Donostia-San Sebastián, 
Basque Country, Spain}
\affiliation{IKERBASQUE, Basque Foundation for Science, 48009 Bilbao, Basque Country, Spain}
\affiliation{Donostia International Physics Center (DIPC), 20018 Donostia--San Sebasti\'an, Spain}

\author{Yao Lu}
\affiliation{
Centro de F\'{i}sica de Materiales (CFM-MPC), Centro Mixto CSIC-UPV/EHU,  E-20018 San Sebasti\'{a}n, Spain}
\author{F. Sebastian Bergeret}
\email{fs.bergeret@csic.es}
\affiliation{
Centro de F\'{i}sica de Materiales (CFM-MPC), Centro Mixto CSIC-UPV/EHU,  E-20018 San Sebasti\'{a}n, Spain}
 \affiliation{Donostia International Physics Center (DIPC), 20018 Donostia--San Sebasti\'an, Spain}

\title{ Spin–galvanic response to non-equilibrium spin injection in superconductors with spin–orbit coupling}

\date{\today}
\begin{abstract}
We show that nonequilibrium spin injection into a superconductor can generate an anomalous supercurrent or induce a phase gradient, even for spin voltages below the superconducting gap.  Our mechanism does not require breaking time-reversal symmetry in the effective superconducting Hamiltonian, but instead relies on nonequilibrium spin injection. We further demonstrate that superconductivity enhances spin injection due to the large quasiparticle density of states near the pairing gap, an effect that persists well below the gap. This contrasts with earlier works predicting the absence of spin injection at zero temperature and small spin voltages. Our results provide a natural explanation for long-standing experimental observations of spin injection in superconductors and predict novel effects arising from spin–charge coupling, including the electrical control of anomalous phase gradients in superconducting systems with spin–orbit coupling. These effects are broadly testable in a variety of materials and hybrid superconducting structures.

\end{abstract}

\maketitle

Recent  research on quantum materials exhibiting superconductivity and spin–orbit coupling, as well as on superconductor–heavy-metal hybrids, has driven growing interest in non-reciprocal transport effects in superconducting systems. \cite{smidman2017superconductivity,nadeem2023superconducting,kokkeler2024nonreciprocal, jeon2022zero,amundsen2024colloquium}. These effects may arise in different materials and structures in which both time-reversal and inversion symmetries are broken. Of particular interest are non-reciprocal effects in superconductors with intrinsic spin–orbit coupling. In this case, the resulting non-reciprocity is closely connected to the superconducting spin-galvanic effect (SGE), namely the conversion of an equilibrium spin polarization—originating from the response of the superconducting condensate to an external magnetic field—into a charge current. 
The SGE takes place in gyrotropic materials or structures, \textit{i.e.}, those that allow for  the existence of a second-rank pseudotensor~\cite{he2020magnetoelectric,kokkeler2025universal,kokkeler2024nonreciprocal}.
Most studies on superconducting SGE have so far focused on supercurrents, {\it i.e.}, on equilibrium properties, where time-reversal symmetry is broken either by an external dc magnetic field or by intrinsic exchange fields, such as those produced by a ferromagnet.

In this work,  we focus on another situation that is also fundamental for realistic experiments: the injection of a non-equilibrium spin density into a superconductor with spin–orbit coupling and its conversion into a voltage drop or a charge current.  The electrical injection of spin into superconductors is a long-standing research topic that probably started with the experiments by Tedrow and Meservey~\cite{tedrow1971spin,tedrow1973spin}, in which the tunneling conductance of ferromagnet-insulator-superconductor (FIS) junctions was used to determine the spin polarization of the F electrodes, and the theoretical work by Aronov~\cite{aronov1976spin}. The basic assumption of the model, sufficient to extract the polarization of the F layers, is that the spectrum of the superconductor remains unaltered. This assumption has propagated throughout the literature since then~\cite{johnson2001spin,takahashi2003spin,morten2004spin,poli2008spin,yang2010extremely,mal2012triplet,quay2013spin,beckmann2016spin,wolf2013spin,bergeret2018colloquium}. 
While this assumption agrees well with experiments on non-local transport over length scales much larger than the superconducting coherence length, especially in spin-split superconductors\cite{quay2013spin,wolf2013spin,heikkila2019thermal}, it fails to describe non-local effects over distances of the order of the superconducting coherence length. As an example, in Ref.~\cite{poli2008spin} spin is injected from a ferromagnetic injector and the non-local resistance is measured at a ferromagnetic detector some distance away. An unmodified BCS spectrum predicts an unbounded increase of the non-local resistance in the superconducting state as $T \!\to\! 0$ \cite{takahashi2003spin}, whereas the experiment clearly shows saturation at low temperatures. Using the kinetic theory for superconductors, we show that such a finite non-local spin signal occurs doe to inevitable renormalization of the spectrum of the superconductor at the injection region, even in the absence of any inelastic relaxation processes. Our theory also naturally explains the change in the effective spin-diffusion length observed in Refs.~\cite{beckmann2004evidence,poli2008spin,gu2002direct,urech_enhanced_2006}. 
Specifically, we show that 
the spin penetration length is set by the minimum of the normal state spin relaxation length $l_S$ and the  superconducting coherence length $\xi_0=\sqrt{D/2\Delta}$. For the injector spin voltage $V_S$ well below the superconducting gap $\Delta$,  the injected spin density grows linearly with $V_S$, and for $V_S\lesssim \Delta$ the spin density injected into the superconductor is larger than in the normal state, resulting in an excess spin for $V_S>\Delta$.

We then examine how the injected spin gives rise to an electrical signal. In a superconductor with spin–orbit coupling, the injected spin naturally couples to charge transport, and can be converted into a measurable electrical response via the SGE. This manifests in different forms, as a voltage, an anomalous phase, or a current, depending on the measurement setup and on the spin voltage at the injector. Here we outline our main results.

 In the open circuit  setup shown in Fig.~1(c), non-equilibrium spin injection into the superconductor through the middle normal-metal finger $N$ induces a charge imbalance when $V_S> \Delta$.  Once generated, the charge imbalance relaxes away from the contact over the characteristic charge-imbalance length $\Lambda^*$ \cite{tinkham1972theory,hubler2010charge}. As a result, a finite voltage is detected between two normal probes, 1 and 2 in Fig.~1(c), located at distances shorter than $\Lambda^*$.  If one probe is placed at a distance much larger than $\Lambda^*$, the voltage difference between this probe and probes~1 or~2 acquires the opposite sign.
In this open-circuit configuration, no net charge current flows through the superconductor; nevertheless, a finite phase difference develops between the two ends of the wire. When the length of the superconducting wire exceeds $\Lambda^*$, the phase difference between the ends of the S wire becomes independent of the wire length. In this sense, spin injection realizes a ``phase battery''~\cite{strambini2020josephson}: a superconducting element that generates a persistent current when embedded in a superconducting loop, as illustrated in Fig.~1(d). The anomalous phase implies a superconducting diode effect in the nonlinear regime \cite{kokkeler2024nonreciprocal}. Specifically, the superconductor in Fig.~1(a) exhibits direction-dependent critical currents when a nonequilibrium spin polarization is induced. Unlike previous proposals, this mechanism does not require time-reversal symmetry breaking in the Hamiltonian but arises from the nonequilibrium spin population.
In the superconducting state, the  loop geometry, Fig.~1(d),  has a  richer phenomenology than in the normal state~\cite{omori2014inverse}.
If the injection spin $V_S<\Delta $, then a non-dissipative current, supercurrent, flows along the loop. In contrast,  if $V_S>\Delta$,  
a  conversion of a quasiparticle (dissipative) charge current into a pure  supercurrent takes place at distances of the order of $\Lambda^*$. Thus,  when the voltage is measured between the two probes located within a distance smaller than $\Lambda^*$ from the injector, upper two  probes in Fig.~1(d),  a finite voltage will be meassured. However, measuring the voltage between probes far from the injector (lower contacts in Fig.~1(d)) yields zero voltage.
In the remainder of the article, we present the theoretical framework underlying these effects and provide quantitative predictions for the proposed experimental configurations. Technical details are given  in the Supplementary Material (SI)\cite{SI}.

{\it Spin injection revisited.} Here we show that  it is possible to inject spin into a superconductor even at spin voltages much smaller than the gap. The reason is that the density of states (DoS) of a superconductor is, strictly speaking, never exactly zero at $\omega <\Delta$. This is, of course, not surprising in real superconductors, where inelastic processes, either intrinsic or due to the electromagnetic environment,  lead to an effective  finite Dynes parameter $\eta$ \cite{pekola2010environment,arutyunov2011spatially}, making the in-gap DoS proportional to $\eta$. However, even for an ideal BCS superconductor with $\eta = 0$, as assumed in previous works \cite{takahashi2003spin,morten_spin_2005,bergeret2018colloquium,heikkila2019thermal}, the {very presence of the normal injector}
leads to a renormalization of the superconducting DoS in the vicinity of the contact. 
The {induced in-gap} DoS $N_\omega$ in the tunnel limit,  is of the order of $\tilde{\lambda}_{\omega}=\frac{\lambda\xi_{\omega}} {D}$, where $\lambda=1/(2e^2R_b N_F)$, $R_b$ is the resistance of the N/S interface per unit area, $\xi_\omega$ the energy dependent superconducting coherent length,  and $N_F$ is the normal DoS  at the Fermi level. 

\begin{figure}[t]
\includegraphics[scale=0.13]{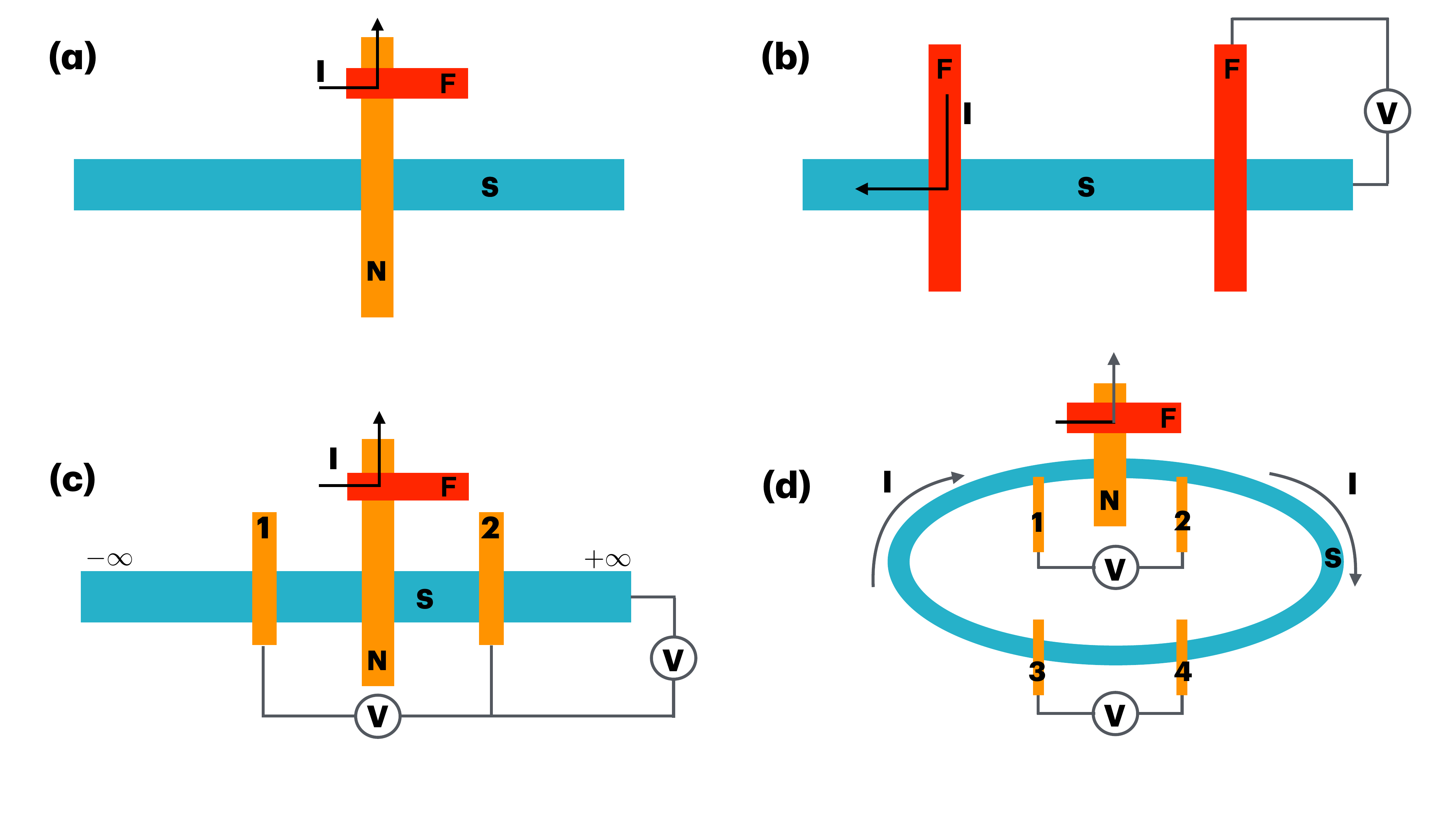}
\caption{ (a)  Schematic  of the system under investigation: A normal metal(N) layer sits atop a superconductor. A spin polarized current is injected from a ferromagnetic lea (F). This results in spin accumulations within the normal metal layer, which  diffuses toward the region of the metal that lies above the superconductor (S). (b)}\label{Fig:Setup}
\end{figure}

To demonstrate this, we consider the setup shown in Fig.~\ref{Fig:Setup}(a), consisting of a quasi-one-dimensional superconducting wire (S) attached to a normal-metal injector at $x=0$. To disentangle spin and charge transport, the electrical spin injection from a ferromagnet F is performed in the $N$ metal, away from the N/S interface. The non-equilibrium spin injected in $N$ diffuses toward the interface and is described by an effective spin voltage $V_S$~\cite{morten2004spin}.
The properties of the superconductor {possessing SGE} are described using the Usadel equation in the Keldysh formalism \cite{virtanen2022nonlinear,kokkeler2025universal}:
\begin{eqnarray}
\label{eq:Usadel_inj}
&-D\partial_{x}(\check{g}\partial_{x}\check{g})+ \partial _x{\cal J}_x^{sg}+[-i\omega\tau_{3}+\Delta\tau_{1},\check{g}]+\nonumber&\\
& +\frac{1}{8\tau_{s}}[\bm{\sigma}\check{g}\bm{\sigma},\check{g}]+\mathcal{T}^{sg}=-\lambda \delta(x)\left[\check{g}_{{\rm inj}},\check{g}\right]\; .&
\end{eqnarray}
Here, $\check{g}$ denotes the quasiclassical Green's function (GF) matrix in the combined Keldysh--Nambu--spin space, $\sigma_j$ and $\tau_j$ are the Pauli matrices acting in spin and Nambu space, respectively, while $\tau_s$ is the spin-relaxation time, which we assume to originate  from  spin-orbit scattering. {Matrix current} ${\cal J}_k^{sg}$ and {torque} $\mathcal{T}^{sg}$ are proportional to the SGE pseudotensor $\gamma$, {and describe the direct and inverse SGE, respectivelly} \cite{kokkeler2025universal}.  Their exact form is given in Eq.~(\ref{J-sg})-(\ref{T-sg}) of SI \cite{SI}.
The right-hand side in Eq.~(\ref{eq:Usadel_inj}) {models a local normal injector characterized by the GF}
$\check{g}_{\mathrm{inj}}$ {with components, $g^R_{\mathrm{inj}}=-g^A_{\mathrm{inj}}=\tau_3$ and $g^K_{\mathrm{inj}}=2\tau_3 f_{\mathrm{inj}}$, where}
\begin{equation}
    f_{\mathrm{inj}} = \tanh\left[{(\omega + \sigma_{y} V_S)}/{2T}\right] 
    = n_{L} + \sigma_{y} n_{T} \, ,
\end{equation}
is the distribution function of the normal electrode {with the applied spin bias $V_S$}. 

Assuming a small $\gamma$, we solve the problem in two steps. {First, we set $\gamma = 0$ and solve the spin-injection problem. Then, the SGE is analyzed perturbatively in $\gamma$.}

From Eq.~(\ref{eq:Usadel_inj}) it follows that the retarded and advanced components contain only singlet terms, $g^{R(A)}$, being proportional to the identity matrix in spin space.
In contrast, the Keldysh component acquires both singlet and triplet contributions, $g^{K} = g_{s}^{K} + \sigma_{y} g_{t}^{K}$.  It is customary \cite{LarkinOvchinnikov1986,morten2004spin,bergeret2018colloquium} to use the parametrization
$g_t^K = \left(g^R - g^A\right) F_t \sigma_y$.   Here $F_t$ is the triplet distribution function which satisfies the following kinetic equation obtained  from Eq. (\ref{eq:Usadel_inj}) (see section \ref{subsec:noDynes} in SI):
\begin{equation}
\label{eq:KE}
-D\partial_x\left[ d_\omega(x)\partial_x F_t\right]+\frac{d_\omega(x)}{\tau_s}F_t=2\lambda\delta(x)N_\omega(0)\left(n_T-F_t\right)
\end{equation}
where  $N_\omega(x)=
 \text{tr} \tau_{3}(g^{R}(x)-g^{A}(x))/4$ is the density of states, and $d_\omega(x)={\rm tr} (1-g^{R}(x)g^A(x))/4$, is a spectral function renormalizing the diffusion coefficient and the spin scattering rate.   In the normal state,   $g^{R} = -g^{A} = \tau_{3}$, and the solution of Eq. (\ref{eq:KE}) takes the form:
\begin{equation}
    F_{t}^N(\omega,x) = \frac{R_{s}/2}{R_{b} + R_{s}/2} 
    e^{-\frac{|x|}{l_{s}}} \, n_{T}(\omega) ,
    \label{F_t-normal}
\end{equation}
where $R_{s} = \rho_{D} l_{s} = l_{s}/(N_{F} D)$ is the 
spin resistance per unit area. 
As expected, the spin injected depends on  the ratio between the barrier and spin resistances,  and 
decays exponentially away from the injector on the {spin diffusion} length scale 
$l_{s}=\sqrt{D\tau_s}$. 

In the superconducting state, assuming the tunneling limit, $R_{b} \gg R_{s}$, the spectral functions $N_\omega (x)$ and $d_\omega(x)$ can be approximated by their values at the injector, $x=0$, multiplied by $\exp(-\tilde\varkappa_\omega|x|)$ {and $\exp(-2\tilde\varkappa_\omega|x|)$}, respectively, where
$
\tilde\varkappa_\omega = \Theta(\Delta^2-\omega^2)[1 - (\omega/\Delta)^2]^{1/4}/\xi_0
$.  {The corresponding solution to Eq.~(\ref{eq:KE}) reads} (SI, Sec.~\ref{subsec:noDynes}):
\begin{equation}
    \label{eq:Ft_gen}
    F_{t}(\omega,x) =
    \frac{\lambda N_{\omega}(0) n_T(\omega)}
         {D d_{\omega}(0)\varkappa_t(\omega) + \lambda N_{\omega}(0)}
     e^{-\varkappa_t(\omega)|x|},
\end{equation}
where $
\varkappa_t = \sqrt{\tilde\varkappa_\omega^2+\varkappa_S^2}-\tilde\varkappa_\omega$ {and $\varkappa_S=1/l_S$.}  This expression smoothly interpolates between different known limits. In particular, {in the normal state,} Eq.~(\ref{eq:Ft_gen}) is recovered after setting $N_\omega=d_\omega=1$, and $\varkappa_t=\varkappa_S$.


\begin{figure}[t!]
\includegraphics[width=0.58\textwidth]{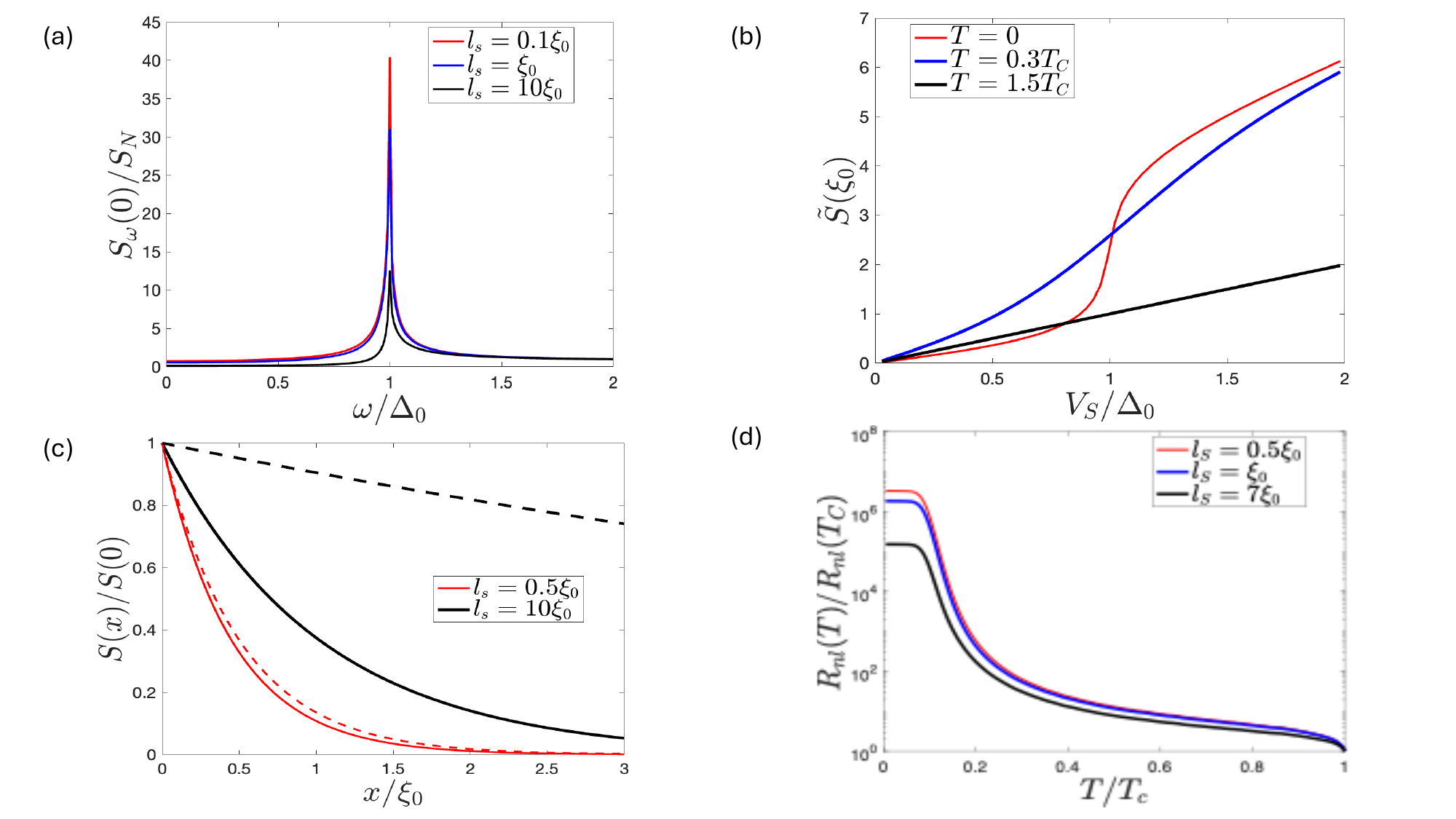}
    \caption{(a) Spectral spin at the injector point, $x=0$ for different values of $l_S$, $T=0$, and $\tilde \lambda=0.1$.  (b)  Dependence of the spin density at $x=\xi_0$ on the spin voltage for different temperatures, $\tilde \lambda=0.05$, and $l_S=\xi_0$. (c) Spatial dependency of the injected spin density in the normal state (dashed lines) and superconducting state (solid lines)  for two different values of $l_S$, $\tilde \lambda=0.05$, $T=0$, and $V_S=0.5\Delta_0$. (d) Temperature dependence of the  non-local resistance measured in the setup of Fig. \ref{Fig:Setup}(b) for different values of $l_S$, and $\tilde\lambda=10^{-3}$. The detector is situated at a distance $2\xi_0$ from the injector.  }
    \label{fig:2}
\end{figure}

Finally,  the spectral spin density is obtained from $S_\omega(x)=\frac{1}{4}{\rm tr}\left\{ \tau_{3}g_{t}^{K}(\omega)\right\} $: 
\begin{equation}
\label{eq:S(x)}
    S_{\omega}(x)=\frac{\lambda N_\omega^2(0) n_{T}(\omega)}{Dd_\omega(0)\varkappa_t(\omega)+\lambda N_\omega(0)}e^{-\sqrt{\tilde\varkappa_\omega^2+\varkappa_S^2}|x|}
\end{equation}
This is an  important result of the first part.   It expresses the non-equilibrium spin density in terms of the spectral functions evaluated at the injection point, $x = 0$. These can be found by solving the algebraic equation, Eq. (\ref{algebraic}) in the SI.
The spectral spin density at $x=0$  is shown in Fig.~\ref{fig:2}a. It is clearly non-zero below the gap. In the limit $l_S \ll \xi_0$, the zero-energy spin density coincides with its normal-state value. Increasing $l_S$ the {subgap} spin density decreases but remains finite {and $\propto\lambda$ as in the normal state. This remarkable result stemms from the fact that even for an ideal BCS superconductor with zero Dynes parameter, the subgap spectral functions remain finite, with $N_\omega(0) \propto \lambda$ and $d_\omega(0) \propto \lambda^2$. When the energy approaches $\Delta$, the spectral spin strongly increaces, having a sharp maximum at $\omega=\Delta$.} 
This behavior contrasts with the assumptions made in previous works \cite{takahashi2003spin,morten2004spin,poli2008spin}, and as we will see has consequences in interpreting real data. 

The spin density is obtained as $S(x)=\frac{N_F}{2}\int d\omega S_\omega(x)$. In Fig.~\ref{fig:2}b we show $S(\xi_0)$, for $l_S = \xi_0$, as a function of the spin voltage. Remarkably, {the maximum in the spectral spin around $\Delta$ translates to an excess spin over the normal state}
for spin voltages exceeding the gap.   In Fig.~\ref{fig:2}c we show the spatial dependence of $S(x)$ at low temperatures (solid lines) and in the normal state (dashed lines). According to Eq.~(\ref{eq:S(x)}), the characteristic spin-diffusion length in the superconducting state is determined by the minimum of $l_S$ and the superconducting coherence length $\xi_0$. In the case $l_S > \xi_0$, as in aluminum, the spin signal decays faster in the superconducting state, {as indeed observed in} Refs.~\cite{beckmann2004evidence,poli2008spin}. In fact,  the ratio of spin diffusion lengths at $T=0$ and {$T>T_c$} gives $\xi_0/l_S$, which using the values for $\Delta$, $D$, and $l_S$ of that reference gives  $\xi_0/l_S\approx 0.10-0.17 $ in  agreement with the experiment. 

From the kinetic equations we can also compute the non-local transport signal in a non-local spin valve, such as the one sketched in Fig. \ref{Fig:Setup}(b), and studied in experiments ~\cite{poli2008spin, beckmann2016spin}.  Details of these calculations are given in Sec~\ref{sec:RNL} of SI.  The obtained non-local resistance $R_{NL}$, {\it i.e.} the ratio between the voltage measured at the detector and the {injector current $I$}, is shown in Fig.~\ref{fig:2}d as a function of temperature. 
We find that, in agreement with the observations \cite{poli2008spin}, $R_{NL}$ does not diverge at $T=0$ as predicted by previous models, but instead saturates as $T \rightarrow 0$, {because of the finite density of states $N_0$} at zero energy.
{The above calculations} assumed a perfect BCS superconductor with vanishing Dynes parameter. The finite {subgap DoS $N_0\sim\lambda\xi_0/D=\tilde{\lambda}$} arises from the tunnel injector, and the zero-temperature value of {$R_{\mathrm{NL}}\sim N_0^{-2}$ is therefore proporional to $1/\tilde\lambda^{2}$.} If inelastic processes dominate ($\eta>\tilde{\lambda}\Delta$), where $\eta$ is the Dynes parameter, the density of states is given by  
$N_\omega={\rm Re} \large[-i(\omega+ i\eta)/\sqrt{\Delta^2-(\omega+ i\eta)^2}\large]$, and 
hence $N_0\sim\eta/\Delta$ controls the low-temperature behavior of $R_{\mathrm{NL}}$.
This is the case of highly resistive tunneling contacts (see Sec.~\ref{sec:Inj_Dynes} of SI), {which we assume in the following}.

{\it Spin-charge conversion.}
Having established the mechanism of spin injection into a superconductor, we now study how the spin is converted into a charge signal via the SGE, for both subgap ($V_S<\Delta$) and above-gap ($V_S>\Delta$) spin biases.
The triplet {GF $\check{g}_t$, generated by the injection, determines the spin-galvanic current $\mathcal{J}^{sg}_x=\gamma\check{g}_t$, which} now acts as a perturbation in the singlet channel, so that from Eq.~(\ref{eq:Usadel_inj}) we obtain:
\begin{eqnarray}\label{eq:KeldyshGF2}
-D\check{g}_{0}\left(\partial_{x}^{2}\check{g}_{1s}-i\partial_{x}^{2}\theta[\tau_{3},\check{g}_{0}]\right)+[-i\omega\tau_{3}+\Delta\tau_{1},\check{g}_{1s}]\nonumber\\
=-\partial_{x}\left(\gamma\check g_t\right).
\end{eqnarray}
 Here $\check g_0$ denotes the unperturbed BCS GF {with finite $\eta$. We assume higly resistive tunneling cotacts with $\eta/\Delta>\tilde\lambda$, so that $\check{g}_0$ is taken spatially independent.} 
Importantly, to ensure charge conservation, we have  introduced the  superconducting phase $\theta$ via $\check{g} \mapsto e^{-i \tau_3 \theta} \check{g} e^{i \tau_3 \theta}$, which is determined from the condition
$\int{d\omega}\,{\rm tr}\left\{ \tau_{2}g^{K}_s(\omega)\right\} =0$. Details of the calculation are presented in Sec.~\ref{sec:s-2-c} of SI \cite{SI}. 
The charge imbalance induced by the SGE follows from $g^{K}_{1s}$ as $Q_q^{*}=\frac{N_F}{8}\int d\omega\,\mathrm{tr}\, g^{K}_{1s}(\omega)$, which determines the potential drop across the injector~\cite{SI}.
\begin{equation}
e\,\Delta\varphi=\gamma\frac{N_F \lambda \tau_s}{D}\int d\omega\, N_\omega\, n_T(\omega).
\end{equation}
Thus, at low temperatures, {and $\eta\rightarrow 0$}, a finite voltage drop appears only if $V_S > \Delta$. In this case, a finite voltage difference $2V_0$ between probes 1 and 2 in Fig.~\ref{Fig:Setup} will be measured, provided that the probes are located at distances smaller than the charge-imbalance length $\Lambda_{Q^*}$. Measuring the voltage between probe 2 and another probe situated at a distance much larger than $\Lambda_{Q^*}$ yields $V_0$, whereas performing the same measurement on the opposite side gives $-V_0$.
\begin{figure}[t]
  \includegraphics[width=0.4\textwidth]{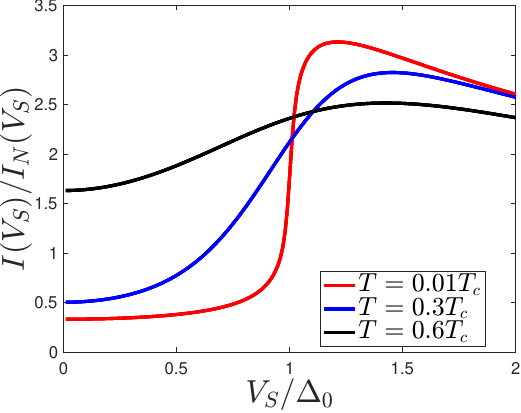}\caption{ The spin voltage dependence of the ratio between the current induced in the superconducting loop of Fig. \ref{Fig:Setup}(d) and its value in the normal state for $l_S=L=\xi_0$, and $\eta=0.01\Delta_0$.
    }
    \label{fig:3}
\end{figure}

The appearance of a charge imbalance in the superconductor is not the only consequence of spin injection. A phase gradient {is also always} generated \cite{SI}. 
We distinguish two experimentally relevant situations. Consider first a superconductor with open boundaries, Fig\ref{Fig:Setup}(c). In this case, no current flows, but a phase gradient is generated to compensate for the SG anomalous current.  The {phase difference $\delta \theta$ between the ends of the wire is found by integrating the continuity equations (see SI, Sec.~\ref{sub-section:phase-drop}) and is given by } $\delta\theta=\gamma\bar S/K_s$,
where $\bar S$ is the total, space integrated, spin induced in the S wire, and $K_s = 2\pi\sigma_0\Delta \tanh(\Delta/2T)$.  
Thus, the finite wire acts as a \textit{phase battery}, {a circuit element} that develops a finite phase difference between its ends. When embedded in a superconducting loop, it drives a circulating current.

This brings us to the second setup, a 
superconducting loop, shown in Fig.~\ref{Fig:Setup}(d). The SGE induces a circulating current given by \cite{SI}
 $I = {\gamma}\bar S/{L} $, 
where $L$ is the loop length. This result  is identical to that of the  normal state.  However, 
in the superconducting state, for large enough values of $V_S$,  $\bar S$ is larger as in the normal state and so the circulating  current. The ratio of the currents in the superconducting and normal state as a function of $V_S$ is shown in Fig. \ref{fig:3}.  If the spin voltage $V_S$ is smaller than $\Delta$, the  current in the loop is a pure supercurrent. In this case, both pairs of probes in Fig.~\ref{Fig:Setup}d will measure zero voltage. In contrast, for $V_S > \Delta$, probes located at distances shorter than $\Lambda_{Q^*}$,  probes~1 and~2 in Fig.~\ref{Fig:Setup}(d), detect a finite voltage, signaling the presence of a dissipative quasiparticle current. At larger distances from the injector, this current is converted into a pure supercurrent. Consequently, probes~3 and~4 in Fig.~\ref{Fig:Setup}(d), which are placed far from the injector, measure zero voltage, indicating purely superconducting transport. In principle, the circulating current could also be detected inductively by a second superconducting loop.

{\it Conclusion}
We have presented a complete theory of the spin-galvanic effect in superconductors induced by a non-equilibrium spin. To this end, we revised the theory of spin injection, showing that even for spin voltages below the superconducting gap a finite spin can be injected. In the presence of intrinsic spin-orbit coupling, this spin generates a charge signal via the SGE.   Specifically, for $V_S < \Delta$, a phase gradient develops, either as a phase drop across the superconductor in an open circuit or as a circulating supercurrent in a loop. For $V_S > \Delta$, a charge imbalance appears, producing a voltage drop across the injector if probes are within the charge-imbalance relaxation length. Our predictions can be tested in standard multiterminal setups used in spintronics, such as nonlocal spin valves~\cite{poli2008spin,beckmann2016spin,jedema2001electrical} and loop geometries~\cite{omori2014inverse}.

{\it Acknowledgments}
We thank financial support from the Spanish MCIN/AEI/10.13039/501100011033 
through the grants PID2023-148225NB-C31, and PID2023-148225NB-C32, and   from the European Union’s Horizon Europe research and innovation program under grant agreement No. 101130224 (JOSEPHINE).

\newpage

\bibliography{refs}

\newpage 

\onecolumngrid
\appendix

\section{ Usadel equation for a superconductor with spin-orbit coupling}

 The complete Usadel equation for a superconductor with intrinsic spin-orbit coupling reads \cite{kokkeler2025universal}:
\begin{equation}
-D\partial_{k}(\check{g}\partial_{k}\check{g})+\partial_{k}\mathcal{J}_{k}^{sg}+[-i\omega\tau_{3}+\Delta\tau_{1}+\check{\Sigma},\check{g}]+\frac{1}{8\tau_{s}}[\bm{\sigma}\check{g}\bm{\sigma},\check{g}]+\mathcal{T}^{sg}=0\; . 
\label{SGE-Usadel-gen}
\end{equation}
Here $\tau_s$ is the spin diffusion time, $D$ the diffusion coefficient, $\omega$ real frequency, and $\Delta$, the superconducting order parameter.
The SGE shows up in the  matrix current,$\mathcal{J}_{k}^{sg}$, which transform the triplet GF into a singlet current
\begin{equation}
\mathcal{J}_{k}^{sg}=\frac{i}{16}\epsilon_{ijl}\gamma_{lk}\left\{ \left[\check{g},\sigma_{i}\right],\sigma_{j}+\check{g}\sigma_{j}\check{g}\right\} \mapsto\gamma_{jk}\check{g}_{t,j}\; ,\label{J-sg}
\end{equation}
where $\gamma$ is the SGE pseudotensor, and  the SGE matrix torque, $\mathcal{T}^{sg}$,  transforms a singlet current into
spin torque,
\begin{equation}
\mathcal{T}^{sg}=\frac{i}{8}\epsilon_{ijl}\gamma_{lk}\left[\left\{ \partial_{k}\check{g},\check{g}\sigma_{i}\check{g}\right\} ,\sigma_{j}\right]\label{T-sg}
\end{equation}
{
The Usadel equation is complemented by the normalization condition $\check g^2=1$  and appropriate boundary conditions. We consider a quasi-one-dimensional superconducting wire (S), shown in Fig.~1(a), in contact with a normal-metal injector (N) on top. The S/N interface is described by the Kupriyanov--Lukichev boundary condition \cite{kuprianov1988influence}
\begin{equation}
n_k\,\check g\,\partial_k \check g
=
\lambda(x)\left[\check g_{\rm inj},\check g\right]\; ,
\end{equation}
where $n_k$ is the unit vector normal to the interface and $\lambda(x)$ is a parameter proportional to the inverse interface resistance. Since the wire is quasi-one-dimensional and the S/N interface cross section is assumed to be smaller than the spin-diffusion length, we approximate $\lambda(x)$ by a delta function, $\lambda(x)=\lambda\,\delta(x)$, with $\lambda=1/(2e^2 R_b N_F)$. In this case the problem reduces to a one dimensional  problem, and the boundary problem is reduced  to  Eq.~(\ref{eq:Usadel_inj}) of the main text.
}

In what follows we focus on the  spin injection from a local
contact with a given spin voltage, $V_S$,  into a quasi-1d system along $x$--direction. Assuming a small spin--galvanic coefficient $\gamma$,  the SGE problem is solved in two steps: (i) the spin injection at $\gamma_{jk}=0$ (section \ref{sec:spin-injection}, and (ii) the spin-charge conversion  perturbatively in $\gamma_{jk}$ (section \ref{sec:s-2-c}). In what follows, we use the $\check .$ for denoting 8$\times$8 matrices, Keldysh-Nambu-spin space. The $\hat .$ symbol denotes 4$\times$4 in the Nambu-spin space, whereas $g$'s denote 2$\times$2 in the Nambu space.

\section{Spin injection problem: Spin-biased injector}

Here we focus on the spin-injection problem neglecting first the SGE. The Usadel equation reads:
\begin{equation}
-D\partial_{k}(\check{g}\partial_{k}\check{g})+[-i\omega\tau_{3}+\Delta\tau_{1},\check{g}]+\frac{1}{8\tau_{s}}[\bm{\sigma}\check{g}\bm{\sigma},\check{g}]+\lambda(x)\left[\check{g}_{{\rm inj}},\check{g}\right]=0\label{Usadel-1}
\end{equation}
where $\lambda(x)$ is a tunneling rate from the injector  assumed
to be localized at the origin.   The normal metal injector, N finger in Fig. \ref{Fig:Setup}(a) in the main text, is described 
by following GFs: 

\begin{align}
g_{{\rm inj}}^{R} & =-g_{{\rm inj}}^{A}=\tau_{3},\label{g-R-injector}\\
\hat g_{{\rm inj}}^{K} & =2\tau_{3}\tanh\left(\frac{\omega+\sigma_{y}V_{s}}{2T}\right)=2\tau_{3}\left(n_{L}+\sigma_{y}n_{T}\right)\; , \label{g-K-injector}
\end{align}
where $V_{s}$ is an effective  ``spin bias'', induced, for example, by the electrical spin injection from a ferromagnet (see \ref{Fig:Setup}(a) in the main text).

Equations for the retarder/advanced components read
\begin{equation}
-D\partial_{x}(g^{R,A}\partial_{x}g^{R,A})+\left[-i\omega\tau_{3}+\Delta\tau_{1},g^{R,A}\right]\pm\lambda(x)\left[\tau_{3},g^{R,A}\right]=0\label{R-Usadel}
\end{equation}
In fact,  the spectral GF have only singlet (scalar)  component. In contrast,
the Keldysh component acquires both singlet and triplet parts, 
\begin{equation}
\hat g^{K}=g_{s}^{K}+\sigma_{y}g_{t}^{K}\; .
    \end{equation}
The triplet part satisfies the 
\begin{align}
-D\partial_{x}\left(g^{R}\partial_{x}g_{t}^{K}+g_{t}^{K}\partial_{x}g^{A}\right)+\left[-i\omega\tau_{3}+\Delta\tau_{1},g_{t}^{K}\right] & +\frac{1}{2\tau_{s}}\left(g^{R}g_{t}^{K}-g_{t}^{K}g^{A}\right)\nonumber \\
+\lambda(x)\left(\tau_{3}g^{K}+g^{K}\tau_{3}-g^{R}2\tau_{3}n_{T}+2\tau_{3}n_{T}g^{A}\right) & =0\label{K-t-Usadel}
\end{align}
In order to obtain the injected spectral spin density, $S_\omega(x)=\frac{1}{4}{\rm tr}\left\{ \tau_{3}g_{t}^{K}(\omega)\right\} $ we will  solve 
Eqs. (\ref{R-Usadel}-\ref{K-t-Usadel}) 
in different situations. From the spectral spin one obtains the physical spin from $S=(N_F/2)\int d\omega S_\omega$.

\subsection{Spin injection in the  normal state}\label{sec:spin-injection}

In the normal state $\Delta=0$ and $g^{R}=-g^{A}=\tau_{3}$, which
dramatically simplifies the problem.  It follows from
\eqref{K-t-Usadel} that $g_{t}^{K}=2\tau_{3}F_{t}$, where the distribution
function satisfies the equation
\begin{equation}
-D\partial_{x}^{2}F_{t}+\frac{1}{\tau_{s}}F_{t}=2\lambda\delta(x)\left(n_{T}-F_{t}\right)\label{KE-normal}
\end{equation}
 where I assumed that the size of the injector is much smaller than
the spin diffusion length $l_{s}=\sqrt{D\tau_{s}}$, so that the local
tunneling rate can be represented as $\lambda(x)=\lambda\delta(x)$, with $\lambda=1/2e^2R_bN_F$, and $R_{b}$ is the contact (barrier) resistance
per area,
The solution of  \eqref{KE-normal} is:
\begin{equation}
F_{t}(\omega,x)=\frac{\lambda l_{s}/D}{1+\lambda l_{s}/D}e^{-\frac{|x|}{l_{s}}}n_{T}(\omega)=\frac{R_{s}/2}{R_{b}+R_{s}/2}e^{-\frac{|x|}{l_{s}}}n_{T}(\omega)\; .\label{F_t-normal}
\end{equation}
where  $R_{s}=l_{s}/\sigma_D$ is the  spin resistance per area, with $\sigma_D$ being the Drude conductivity $\sigma_D=e^2N_F D$. {The factor of 1/2 in Eq.~\eqref{F_t-normal} signifies the fact that the parts of the wire on the opposite sides from the injector work effectively as parallel resistors.} Thus,  the injection in the normal {metallic wire} is characterized
by a single dimensionless parameter,
\[
\frac{\lambda l_{s}}{D}=\frac{\lambda N_{F}l_{s}}{N_{F}D}=\frac{R_{s}}{2R_{b}}
\]
that is the ratio of characteristic resistances. The tunneling limit corresponds to the regime  $R_{b}>>R_{s}$.
 In the opposite limit of $\lambda l_{s}/D>>1$ the contact
is transparent. In what follows, we focus on    tunneling contacts. 
\subsection{Spin injection in ideal superconductors with vanishing Dynes parameters}\label{subsec:noDynes}

We first focus on a perfect BCS superconductor without inelastic processes. This situation conicides with previous theory works\cite{takahashi2003spin,morten2004spin,morten_spin_2005,heikkila2019thermal,silaev2015long}.
It is customary to parametrize the Keldysh GF $g_{t}^{K}$ as\cite{bergeret2018colloquium},
\begin{equation}
g_{t}^{K}=(g^{R}-g^{A})F_{t}\label{F_t-def}
\end{equation}
 where $g^{R,A}$ are the solutions to \eqref{R-Usadel}. By inserting
the representation \eqref{F_t-def} into Eq.\eqref{K-t-Usadel}, and
taking its trace we obtain  the following exact kinetic equation for
the distribution function in the SC
\begin{align}
-D\partial_{x}\left(\text{tr}\left\{ 1-g^{R}g^{A}\right\} \partial_{x}F_{t}\right)+\frac{1}{\tau_{s}}\text{tr}\left\{ 1-g^{R}g^{A}\right\} F_{t}\nonumber \\
=2\lambda(x)\text{tr}\left\{ \tau_{3}\left(g^{R}-g^{A}\right)\right\} \left(n_{T}-F_{t}\right)\label{KE-SC}
\end{align}
It has a structure similar to Eq.\eqref{KE-normal} in
the normal state, but with renormalized kinetic coefficients,
\begin{align}
D & \mapsto D_{\omega}(x)=\frac{D}{4}{\rm tr}\left\{ 1-g^{R}g^{A}\right\} =\frac{D}{8}{\rm tr}\left\{ \left(g^{R}-g^{A}\right)^{2}\right\} \equiv Dd_{\omega}(x)\label{d-omega}\\
\frac{1}{\tau_{s}} & \mapsto\frac{1}{\tau_{s}(\omega,x)}=\frac{1}{4\tau_{s}}{\rm tr}\left\{ 1-g^{R}g^{A}\right\} =\frac{1}{8\tau_{s}}{\rm tr}\left\{ \left(g^{R}-g^{A}\right)^{2}\right\} =\frac{d_{\omega}(x)}{\tau_{s}}\nonumber \\
\lambda & \mapsto\lambda_{\omega}(x)=\frac{\lambda}{4}\text{tr}\left\{ \tau_{3}\left(g^{R}-g^{A}\right)\right\} =\lambda N_{\omega}(x)\label{lambda-omega}
\end{align}
In the tunneling regime, the kinetic equation can be solved analytically,
in two limiting cases.
\begin{enumerate}
\item It the limit of small energies $\omega<<\Delta$ a complete analytic
solution is possible because in this case the spatial dependence of
the renormalized kinetic coefficients is exponential.
\item In the limit of a short spin relaxation length, when $l_{s}<<\xi=\sqrt{\frac{D}{2\Delta}}$.
As $d_{\omega}(x)$ and $N_{\omega}(x)$ are varying on the scale
large than $\xi$, while the characteristic scale of eq.\eqref{KE-SC}
is $l_{s}$, we can the spatial dependence of the kinetic coefficient
in \eqref{KE-SC} and solve it the same way as in the normal metal.

\end{enumerate}
In the next two subsections, we present these two cases, which will help to construct a general solution by interpolation. 

\subsubsection{Spin injection in the small energy limit.}

In the case of small energy, and within the tunneling limit, the Usadel
equation for $g^{R,A}(\omega,x),$ eq.\eqref{R-Usadel} can be solved
perturbatively to the linear order in $\lambda\xi/D$. The linearized
in $\lambda$ equation for $g^{R}$ reads
\[
-Dg_{0}^{R}\partial_{x}^{2}g_{1}^{R}+\Omega_{\omega}[g_{0}^{R},g_{1}^{R}]+\lambda(x)[\tau_{3},g_{0}^{R}]=0
\]
where $g_{0}^{R}$ is the BCS GF and $\Omega_{\omega}=\sqrt{\Delta^{2}-\omega^{2}}$.
After multiplication with $g_{0}^{R}$ the above equation takes the
form,
\begin{equation}
-D\partial_{x}^{2}g_{1}^{R}+2\Omega_{\omega}g_{1}^{R}=\lambda(x)\left(\tau_{3}-g_{0}^{R}\tau_{3}g_{0}^{R}\right)\label{eq-g1-R}
\end{equation}
For the kinetic coefficients in Eqs.\eqref{d-omega}-\eqref{lambda-omega}
we need to obtain $g^{R}-g^{A}$. The equation for this difference is obtained
by  subtracting from \eqref{eq-g1-R} its advanced counterpart,
\begin{equation}
-\partial_{x}^{2}\left(g^{R}-g^{A}\right)+\varkappa_{\omega}^{2}\left(g^{R}-g^{A}\right)=\frac{2\lambda}{D}\delta(x)\frac{2\Delta^{2}}{\Delta^{2}-\omega^{2}}\tau_{3}\label{gR-gA-equation}
\end{equation}
where $\varkappa_{\omega}=\xi_{\omega}^{-1}=\sqrt{\frac{2\Omega_{\omega}}{D}}$
is the inverse spectral  coherence length. We have used the fact that below the
gap $g_{0}^{R}=g_{0}^{A}=\frac{-i\omega\tau_{3}+\Delta\tau_{1}}{\Omega_{\omega}}$.
The solution to Eq. (\ref{gR-gA-equation}) reads,
\begin{equation}
g^{R}-g^{A}=2\tau_{3}\frac{\lambda\xi_{\omega}}{D}\frac{\Delta^{2}}{\Delta^{2}-\omega^{2}}e^{-\varkappa_{\omega}|x|}=2\tau_{3}\tilde{\lambda}_{\omega}e^{-\varkappa_{\omega}|x|}\label{gR-gA-solution}
\end{equation}
The validity of this solution assumes the condition
\[
\tilde{\lambda}_{\omega}=\frac{\lambda\xi_{\omega}}{D}\frac{\Delta^{2}}{\Delta^{2}-\omega^{2}}<<1
\]
which is definitely satisfied for low energies in the tunneling regime
when $\tilde{\lambda}_{\omega}\approx\tilde{\lambda}_{0}=\frac{\lambda\xi_{0}}{D}$.
From eq.\eqref{gR-gA-solution} we immediately find the required kinetic
coefficients,
\begin{align}
N_{\omega}(x) & =\frac{1}{4}{\rm tr}\left\{ \tau_{3}\left(g^{R}-g^{A}\right)\right\} =\tilde{\lambda}_{\omega}e^{-\varkappa_{\omega}|x|}\label{N(x)}\\
d_{\omega}(x) & =\frac{1}{8}{\rm tr}\left\{ \left(g^{R}-g^{A}\right)^{2}\right\} =\tilde{\lambda}_{\omega}^{2}e^{-2\varkappa_{\omega}|x|}\label{d(x)}
\end{align}
The kinetic equation \eqref{KE-SC} then takes the following form
\begin{equation}
-e^{2\varkappa_{\omega}|x|}\partial_{x}\left(e^{-2\varkappa_{\omega}|x|}\partial_{x}F_{t}\right)+\varkappa_{s}^{2}F_{t}=\frac{2\lambda}{D\tilde{\lambda}_{\omega}}\delta(x)\left(n_{T}-F_{t}\right)\label{KE-SC1}
\end{equation}
where $\varkappa_{s}=\xi_{s}^{-1}=1/\sqrt{D\tau_{s}}$. It is remarkable
that the r.h.s. in \eqref{KE-SC1} is independent of the tunneling
rate $\lambda$:
\[
\frac{\lambda}{D\tilde{\lambda}_{\omega}}=\varkappa_{\omega}\left(1-\frac{\omega^{2}}{\Delta^{2}}\right)\; .
\]
In other words, in the parametrization Eq. (\ref{F_t-def}), the distribution function does not depend on $\lambda$. 
One can check by a direct substitution that the solution of \eqref{KE-SC1}
is of the form
\[
F_{t}(x)=F_{t}(0)e^{-(\sqrt{\varkappa_{\omega}^{2}+\varkappa_{s}^{2}}-\varkappa_{\omega})|x|}
\]
where the value of $F_{t}(0)$ is found from the boundary condition
\[
\left[\partial_{x}F_{t}\right]_{-0}^{+0}=2\varkappa_{\omega}\left(1-\frac{\omega^{2}}{\Delta^{2}}\right)\left[n_{T}-F_{t}(0)\right]
\]
By resolving this condition we find the final distribution function,
\begin{equation}
F_{t}(x)=\frac{\varkappa_{\omega}\left(1-\omega^{2}/\Delta^{2}\right)n_{T}(\omega)}{\sqrt{\varkappa_{\omega}^{2}+\varkappa_{s}^{2}}-\varkappa_{\omega}\omega^{2}/\Delta^{2}}e^{-(\sqrt{\varkappa_{\omega}^{2}+\varkappa_{s}^{2}}-\varkappa_{\omega})|x|}\label{F(x)}\; . 
\end{equation}
We then obtain for the spectral spin density,
\begin{equation}
S_{\omega}(x)=F_{t}(\omega,x)N_{\omega}(x)=\frac{\lambda n_{T}(\omega)/D}{\sqrt{\varkappa_{\omega}^{2}+\varkappa_{s}^{2}}-\varkappa_{\omega}\omega^{2}/\Delta^{2}}e^{-\sqrt{\varkappa_{\omega}^{2}+\varkappa_{s}^{2}}|x|}
\end{equation}
In the small energy limit this result simplifies as,
\begin{equation}
S_{\omega}(x)\approx\frac{R_{s}}{R_{b}}n_{T}(\omega)\frac{\varkappa_{s}}{\sqrt{\varkappa_{0}^{2}+\varkappa_{s}^{2}}}e^{-\sqrt{\varkappa_{0}^{2}+\varkappa_{s}^{2}}|x|}\label{S-small-E}
\end{equation}
 In the limit of short spin diffusion length $\varkappa_{s}>>\varkappa_{\omega}$
the above small energy spin density in SC coincides with the spectral
spin in the normal metal ({\it cf.} with Eq. (\ref{F_t-normal}) when $R_b\gg R_s$),
\[
S_{\omega}^{N}(x)\approx\frac{R_{s}}{R_{b}}n_{T}(\omega)e^{-\varkappa_{s}|x|}
\]
In short, even in the case of an ideal BCS superconductor with $eta=0$,  the tunneling contact provides a finite density of states at $\omega=0$, which leads to a finite subgap spin density, in contrast with teh assumptions of previous works \cite{takahashi2003spin,morten_spin_2005, poli2008spin}.

\subsubsection{Spin injection in the limit of a short spin diffusion length}

Now, let us assume that the spin diffusion length is short, $l_{s}<<\xi_{0}(0)$,
where $\xi_{0}(0)=\xi_{\omega=0}(T=0)$. In this limit, the spatial
dependence of the kinetic coefficients can be neglected and in \eqref{KE-SC}
we can simply replace $d_{\omega}(x)$ by its value at the injector,
i.e., $d_{\omega}(x)\mapsto d_{\omega}(0)$. Equation \eqref{KE-SC}
then simplifies to the form
\[
-\partial_{x}^{2}F_{t}+\varkappa_{s}^{2}F_{t}=\frac{2\lambda N_{\omega}(0)}{Dd_{\omega}(0)}\delta(x)\left(n_{T}-F_{t}\right)
\]
which is essentially identical to the kinetic equation in the normal
metal and is solved in the same way. The solution is,
\begin{equation}
F_{t}(\omega,x)=\frac{\frac{\lambda l_{s}N_{\omega}(0)}{Dd_{\omega}(0)}}{1+\frac{\lambda l_{s}N_{\omega}(0)}{Dd_{\omega}(0)}}n_{T}(\omega)\,e^{-\varkappa_{s}|x|}\label{F(x)-short-ls}
\end{equation}
Notice that  in the tunneling limit  when $\lambda l_{s}/D=R_{s}/R_{b}<<1$.
The functions $N_{\omega}(0)$ and $d_{\omega}(0)$ are expressed
in terms of components of $g_{\omega}^{R}(x)=g_{\omega}(x)\tau_{3}+f_{\omega}(x)\tau_{1}$
at the injector point,
\begin{align*}
N_{\omega}(0) & =\frac{1}{4}{\rm tr}\left\{ \tau_{3}\left(g^{R}(0)-g^{A}(0)\right)\right\} =\Re g_{\omega}(0)\\
d_{\omega}(0) & =\frac{1}{8}{\rm tr}\left\{ \left(g^{R}(0)-g^{A}(0)\right)^{2}\right\} =\left[\Re g_{\omega}(0)\right]^{2}-\left[\Im f_{\omega}(0)\right]^{2}
\end{align*}
Calculation of $g(0)$ and $f(0)$ can be reduced to solving an algebraic
quartic equation. To derive this equation I represent the Usadel equation
for $g^{R}(x)=g(x)\tau_3+f(x)\tau_1$ in the following explicit form
\begin{align}
-\frac{D}{2}\partial_{x}\left[g(x)\partial_{x}f(x)-f(x)\partial_{x}g(x)\right]-i\omega f(x)-\Delta g(x) & =0\label{R-Usadel-2}\\
-\frac{D}{2}\left[g(x)\partial_{x}f(x)-f(x)\partial_{x}g(x)\right]_{-0}^{+0}=\lambda f(0)\label{R-BC}
\end{align}
Here the boundary condition \eqref{R-BC} represents the injection
$\delta$-function. In addition, the normalization $g^{2}(x)+f^{2}(x)=1$ is
assumed.  By multiplying Eq.\eqref{R-Usadel-2} with $g(x)\partial_{x}f(x)-f(x)\partial_{x}g(x)$
one can represent it in a total derivative form
\[
\partial_{x}\left(\frac{D}{4}\left[g(x)\partial_{x}f(x)-f(x)\partial_{x}g(x)\right]^{2}-i\omega g(x)+\Delta f(x)\right)=0
\]
The integration of this equation from a given point $x$ to $\infty$
gives the first integral of the Usadel equation:
\begin{equation}
\frac{D}{4}\left[g(x)\partial_{x}f(x)-f(x)\partial_{x}g(x)\right]^{2}-i\omega g(x)+\Delta f(x)=\sqrt{\Delta^{2}-(\omega+i0)^{2}}\label{1st integral}
\end{equation}
where the r.h.s. corresponds to $\left[-i\omega g(x)+\Delta f(x)\right]_{x=\infty}$.
We now evaluate eq.\eqref{1st integral} at $x=0$ and substitute
the boundary condition \eqref{R-BC} for the first term. This yields an algebraic
equation,
\[
\frac{\lambda^{2}}{4D}f^{2}(0)-i\omega g(0)+\Delta f(0)=\sqrt{\Delta^{2}-(\omega+i0)^{2}}
\]
which, together with the normalization condition $g^{2}(0)+f^{2}(0)=1$,
fully determines the required functions $g_{\omega}(0)$ and $f_{\omega}(0).$
It is useful to  rewrite the above equation in terms of
dimensionless parameters, 
\begin{equation}
\frac{1}{2}\tilde{\lambda}f^{2}-izg+f-\sqrt{1-z^{2}}=0\label{algebraic}
\end{equation}
where $z=\frac{\omega}{\Delta}+i0$, and $\tilde{\lambda}=\frac{\lambda}{\sqrt{2\Delta D}}=\frac{\lambda\xi_{0}}{D}$
is the only dimensionless parameter which controls the solution. By
squaring this equation can be reduced to a closed quartic equation
either for $g$ or for $f$. It becomes quadratic at $z=0$, and also
possesses a relatively simple analytic solution in the gap region
at $z\approx1$ in the limit $\tilde{\lambda}<<1$. Specifically,
assuming $\tilde{\lambda}<<1$, and solving the squared equation for
$g$ we obtain for zero energy,
\[
g_0(0)\approx\tilde{\lambda}
\]
which agrees with Eq.\eqref{gR-gA-solution}. In the
vicinity of the gap for $|z-1|<<1$ the solution reads,
\[
g(z)\approx\tilde{\lambda}^{-\frac{2}{3}}\left\{ e^{-i\frac{\pi}{6}}-\frac{2}{3}\tilde{\lambda}^{-\frac{2}{3}}e^{i\frac{\pi}{6}}\sqrt{1-z^{2}}\right\} 
\]
 Notice that since $z=\frac{\omega}{\Delta}+i0$, the square root
in the above equation  reads,
\[
\sqrt{1-z^{2}}=\sqrt{1-z^{2}}\,\theta(1-z^{2})-i\,\sqrt{z^{2}-1}\,\theta(z^{2}-1)
\]

\subsubsection{Interpolation scheme smoothly connecting known limits}

From the previous sections it becomes clear that (i) the tunneling contact modifies the spectral Green's functions, and (ii) these modifications decay on the scale of $\varkappa_{\omega}$,  for energies inside the gap.
Therefore, it looks reasonable to approximate
the space dependence of spectral coefficients $d_{\omega}(x)$ and
$N_{\omega}(x)$, Eqs. \eqref{d-omega} and \eqref{lambda-omega},
by the following simple exponential functions,
\begin{align}
N_{\omega}(x) & =N_{\omega}(0)e^{-\bar{\varkappa}_{\omega}|x|}\label{N-approx}\\
d_{\omega}(x) & =d_{\omega}(0)e^{-2\bar{\varkappa}_{\omega}|x|}\label{d-approx}
\end{align}
Here $N_{\omega}(0)$and $d_{\omega}(0)$ are the exact values at
the contact, and we define $\bar{\varkappa}_{\omega}^{2}=\theta(\Delta^{2}-\omega^{2})2\sqrt{\Delta^{2}-\omega^{2}}/D$
which takes care of the fact that above the gap the density of states
is well approximated by a space independent function. In fact, in the limit of small
tunneling rate $\lambda$ the essential relative modifications of
all the spectral coefficients are in the gap region.   Note that the heuristic construction of Eqs.~\eqref{N-approx}-\eqref{d-approx} is inspired by the small-energy solution of Eqs.~\eqref{N(x)}-\eqref{d(x)}. With this assumption,  the kinetic equation, Eq.  \eqref{KE-SC}, takes the form structurally identical  to its low energy form
\eqref{KE-SC1},
\begin{equation}
-e^{2\bar{\varkappa}_{\omega}|x|}\partial_{x}\left(e^{-2\bar{\varkappa}_{\omega}|x|}\partial_{x}F_{t}\right)+\varkappa_{s}^{2}F_{t}=\frac{2\lambda N_{\omega}(0)}{Dd_{\omega}(0)}\delta(x)\left(n_{T}-F_{t}\right)\label{KE-SC2}
\end{equation}
Therefore the solution of \eqref{KE-SC2} is again of the form
\begin{equation}
F_{t}(x)=F_{t}(0)e^{-(\sqrt{\bar{\varkappa}_{\omega}^{2}+\varkappa_{s}^{2}}-\bar{\varkappa}_{\omega})|x|}\;. 
\end{equation}
 The difference is that now  $F_{t}(0)$ is determined by a slightly modified boundary
condition
\[
\left[\partial_{x}F_{t}\right]_{-0}^{+0}=-\frac{2\lambda N_{\omega}(0)}{Dd_{\omega}(0)}\left[n_{T}-F_{t}(0)\right]
\]
The final solution for the distribution function then takes the form
\begin{equation}
F_{t}(x)=\frac{\lambda N_{\omega}(0)n_{T}}{Dd_{\omega}(0)\varkappa_{t}(\omega)+\lambda N_{\omega}(0)}e^{-\varkappa_{t}(\omega)|x|}\label{F(x)-approx}
\end{equation}
where $\varkappa_{t}(\omega)=\sqrt{\bar{\varkappa}_{\omega}^{2}+\varkappa_{s}^{2}}-\bar{\varkappa}_{\omega}$.
This result for the distribution function,  interpolates smoothly between all previous    limiting cases. Namely, 
\begin{itemize}
\item the short $l_{s}$ limit of Eq. \eqref{F(x)-short-ls}, which corresponds
to $\varkappa_{s}\gg\varkappa_{\omega}$;
\item the low energy limit of Eq. \eqref{F(x)};
\item the normal state limit in which $\Delta=0$ leading to $N_{\omega}=d_{\omega}=1$
and $\varkappa_{t}(\omega)=\varkappa_{s}$. In this limit we recover
Eq. \eqref{F_t-normal}
\end{itemize}
The corresponding spectral spin is therefore 
\begin{equation}
S_{\omega}(x)=F_{t}(x)N_{\omega}(x)=\frac{\lambda N_{\omega}^{2}(0)n_{T}}{Dd_{\omega}(0)\varkappa_{t}(\omega)+\lambda N_{\omega}(0)}e^{-\sqrt{\bar{\varkappa}_{\omega}^{2}+\varkappa_{s}^{2}}|x|}\label{S-omega-approx}
\end{equation}
This expression for the spin also recovers all above limits except
for the case of ``transparent contact'' because for $\lambda\to\infty$
the assumption Eq. \eqref{N-approx} is not valid (in this case gap
closes at the contact point so that $N_{\omega}(0)=1$).

In the main text,  we focus on tunneling contacts. To create Fig. 2 (a-c) we   use Eq.~(\ref{S-omega-approx}) by  solving Eq.~(\ref{algebraic}) to determine the spectral functions $d_\omega(0)$ and $N_\omega(0)$ at the injection point.

\subsection{Spin injection in a  superconductor with a finite
Dynes parameter: tunneling limit}\label{sec:Inj_Dynes}
In the previous sections, we have assumed an ideal BCS superconductor, \textit{i.e.}, a superconductor with an exactly zero density of states below the gap. However, real superconductors do not exhibit a perfect gap, either due to intrinsic inelastic processes or extrinsic ones, such as coupling to the electromagnetic environment~\cite{pekola2010environment}. Such effects may be qualitatively described by a single parameter~$\eta$, the so-called Dynes parameter.
In this section we compute the spin injected in a superconductor with a finite $\eta$. 

We focus  on the tunneling limit and calculate again the Keldysh triplet component. { The retarded and advance GF's are now homogeneous in space}
To the linear order in the tunneling rate the equation has the simple form:
\begin{equation}
-D\partial_{x}^{2}g_t^{K}+(\Omega_{R}+\Omega_{A})g_t^{K}+\frac{1}{\tau_{s}}g_t^{K}=2\lambda \delta(x)(\tau_{3}-g_{0}^{R}\tau_{3}g_{0}^{A})n_{T}
\end{equation}
where $\Omega_{R,A}=\sqrt{\Delta^{2}-(\omega\pm i\eta)^{2}}$ and
the unperturbed GFs are defined as,
\begin{equation}
g_{0}^{R,A}=\frac{-i(\omega\pm i\eta)\tau_{3}+\Delta\tau_{1}}{\Omega_{R,A}}\; .
\label{eq:gRA_0}
\end{equation}
The triplet component of $g^K$ can be obtained straightforwardly:  
\begin{equation}
g_t^K=\frac{\lambda e^{-|x|\sqrt{\tilde{\varkappa}_\omega^2+\varkappa_S^2}}}
{D\sqrt{\tilde{\varkappa}_\omega^2+\varkappa_S^2}}(\tau_{3}-g_{0}^{R}\tau_{3}g_{0}^{A})n_T
\end{equation}.  The spectral spin density, $S_\omega(x)=\frac{1}{4}{\rm tr}\left\{ \tau_{3}g_{t}^{K}(\omega)\right\} $,  is then given by:
\begin{equation} \label{Dynes-spin}
    S_\omega(x)=\frac{\lambda N_\omega^2 n_T(\omega) }{D d_\omega \sqrt{\tilde{\varkappa}_\omega^2+\varkappa_S^2}}e^{-|x|\sqrt{\tilde{\varkappa}_\omega^2+\varkappa_S^2}}
\end{equation}
{ where we have used the following identity for GFs,
\begin{equation} \nonumber
g_{0}^{R}-g_{0}^{A}=\frac{2\eta}{\Omega_{R}+\Omega_{A}}\left(\tau_{3}-g_{0}^{R}\tau_{3}g_{0}^{A}\right)
\end{equation}
By computing the spin density $S(x)$ from the expression of Eq.~\eqref{Dynes-spin}, one can see that it qualitatively behaves like that of an ideal BCS superconductor (Fig.~1b in the main text). First, it does not vanishes for $V_S < \Delta$, and for $V_S\sim \Delta$, the spin density in the superconducting state is much larger than in the normal state.}



\subsection{Electric detection of injected spin by a magnetic probe}

In order to electrically detect the spin injected at $x=0$, one may place a ferromagnetic probe F at a distance $x_d$ from the injector. A sizable spin signal may appear if $x_d$ is smaller than the characteristic decay length of the spin, determined by the minimum of $l_S$ and $xi_0$.

We assume that the  F--electrode, located at the point $x_{d}$,  is polarize in the same direction as the spin injected, {\it i.e.} $\sigma_y$. 
The electrode is described by the following term in the Usadel equation,
\begin{equation}
\check{J}_{d}\delta(x-x_{d})=\lambda_{d}[\Gamma\check{g}_{d}\Gamma,\check{g}]\delta(x-x_{d})\label{J-d}
\end{equation}
where $\lambda_d=1/2e^2N_FR_d$, $R_d$ is the S/F interface resistance times area, and  $\check{g}_{d}$ is the GF of the normal detector electrode
\begin{align}
g_{{\rm d}}^{R} & =-g_{{\rm d}}^{A}=\tau_{3},\label{g-R-detector}\\
g_{{\rm d}}^{K} & =2\tau_{3}\tanh\left(\frac{\omega+\tau_{3}V_{d}}{2T}\right)=2\tau_{3}\left(n_{L}+\tau_{3}n_{T}\right)\label{g-K-detector}
\end{align}
and $\Gamma=t+\tau_{3}\sigma_{y}u$ with $t^{2}+u^{2}=1$ and $2tu=P$
is the spin filtering operator accounting for the spin polarization
of the detector. For a compact characterization of the $F$--electrode
it is natural to introduce the effective GF,
\begin{equation}
\check{G}_{d}=\Gamma\check{g}_{d}\Gamma\label{G-detector}
\end{equation}
whose components are easily found to be,
\begin{align}
G_{d}^{R} & =-G_{d}^{A}=\tau_{3}+P_{d}\sigma_{y}\label{G-R-detector}\\
G_{d}^{K} & =2(\tau_{3}+P\sigma_{y})(n_{L}+\tau_{3}n_{T})\label{G-K-detector}
\end{align}
 The charge current in the detector is calculated as,
\[
I_{d}=-\frac{\pi}{4}N_{F}\int\frac{d\omega}{2\pi}{\rm tr}\left\{ \tau_{3}\check{J}_{d}^{K}\right\} =-\frac{1}{8}N_{F}\lambda_{d}\int d\omega{\rm tr}\left\{ \tau_{3}[\check{G}_{d},\check{g}(x_{d})]^{K}\right\} 
\]
 Using $\check{g}$ generated by the spin--biased electrode we get,
\[
\frac{1}{8}{\rm tr}\left\{ \tau_{3}[\check{G}_{d},\check{g}]^{K}\right\} =N_{\omega}\left(P_{d}F_{t}-n_{T}\right)
\]
Therefore the current in the detector reads 
\begin{equation}
I_{d}=R_{d}^{-1}\frac{1}{2}\int d\omega\left[N_{\omega}(x_{d})n_{T}-P_{d}N_{\omega}(x_{d})F_{t}(x_{d})\right]\label{I-d}
\end{equation}
 The second term in the r.h.s. is the injected spin
at location of the detector times the polarization of the detector.
The condition $I_{d}=0$ determines the detector voltage, which can be obtained by solving following integral equation:
\begin{equation}
\int d\omega N_{\omega}(x_{d})n_{T}(V_{d})=P_{d}\int S_{\omega}(x_{d})d\omega
\end{equation}

\section{Electric spin injection from magnetic electrode}\label{sec:RNL}

In the previous sections, we considered spin injection via a spin voltage induced in a normal (N) electrode; see Fig.~1a in the main text. However, in experiments, spin is usually injected from ferromagnetic (F) electrodes by driving a spin-polarized current \cite{jedema2001electrical,beckmann2004evidence,poli2008spin,beckmann2016spin}. In this section, we analyze this situation.

\subsection{General kinetic equations}\label{sec:generalKE}

The Usadel equation for a wire with ferromagnetic contact at $x=0$
reads,
\begin{equation}
-D\nabla_{x}(\check{g}\nabla_{x}\check{g})+[-i\omega\tau_{3}+\Delta\tau_{1},\check{g}]+\frac{1}{8\tau_{s}}[\bm{\sigma}\check{g}\bm{\sigma},\check{g}]=-\lambda\delta(x)\left[\check{G}_{{\rm F}}(V),\check{g}\right]\label{Usadel-F-general}
\end{equation}
where $\Delta$ is real and $\nabla_{k}g=\partial_{k}g-i[\partial_{k}\varphi\tau_{3},g]$,
which accounts for a possibility of having a supercurrent. The Green
function $\hat{G}_{F}$ of the normal F--electrode with polarization
$P$ along $y$--axis read {[}see Eqs. \eqref{G-detector}-\eqref{G-K-detector}{]},
\begin{align*}
G_{F}^{R} & =-G_{F}^{A}=\tau_{3}+P\sigma_{y}\\
G_{F}^{K} & =2(\tau_{3}+P\sigma_{y})(n_{L}+\tau_{3}n_{T})
\end{align*}
From Eq. (\ref{Usadel-F-general}),   $g^{R,A}$ have trivial spin structure.  The  equation
for them is given by Eq. \eqref{R-Usadel} after replacing $\partial_{k}\to\nabla_{k}$,
that is,
\begin{align}
-D\nabla_{x}(g^{R}\nabla_{x}g^{R})+\left[-i\omega\tau_{3}+\Delta\tau_{1},g^{R}\right]+\lambda\delta(x)\left[\tau_{3},g^{R}\right] & =0\label{g-R-equation}\\
-D\nabla_{x}(g^{A}\nabla_{x}g^{A})+\left[-i\omega\tau_{3}+\Delta\tau_{1},g^{A}\right]-\lambda\delta(x)\left[\tau_{3},g^{A}\right] & =0\label{g-A-equation}
\end{align}
 The Keldysh component is naturally separated into the singlet and
triplet parts,
\[
g^{K}=g_{s}^{K}+\sigma_{y}g_{t}^{K}
\]
and can then be represented in terms of four distribution functions,
\begin{align}
g_{s}^{K} & =(g^{R}-g^{A})F_{s}^{L}+(g^{R}\tau_{3}-\tau_{3}g^{A})F_{s}^{T}\label{g-s-K}\\
g_{t}^{K} & =(g^{R}-g^{A})F_{t}^{T}+(g^{R}\tau_{3}-\tau_{3}g^{A})F_{t}^{L}\label{g-t-K}
\end{align}
{ In previous section, where a pure  spin biased injector was consider,  the triplet channel generates
only $F_{t}^{T}$ and was completely decoupled from the rest. Now however, } by substituting the above representation
for $g^{K}$into the Keldysh component of Eq. \eqref{Usadel-F-general}
and taking the traces with $\tau_{3}$, $1$, $\sigma_{y}$, and $\tau_{3}\sigma_{y}$
one get four coupled equations,
\begin{align}
-D\text{\ensuremath{\partial_{x}\left(\tilde{d}_{\omega}\partial_{x}F_{s}^{T}\right)-Dj_{x,\omega}\partial_{x}F_{s}^{L}}} & +\frac{1}{2}\Delta{\rm tr}\left\{ \tau_{1}(g^{R}+g^{A})\right\} F_{s}^{T}\nonumber \\
 & =2\lambda N_{\omega}\delta(x)\left[n_{T}-F_{s}^{T}-PF_{t}^{T}\right]\label{KE-F-s-T}\\
-D\text{\ensuremath{\partial_{x}\left(d_{\omega}\partial_{x}F_{s}^{L}+j_{x,\omega}F_{s}^{T}\right)}} & =2\lambda N_{\omega}\delta(x)\left[n_{L}-F_{s}^{L}-PF_{t}^{L}\right]\label{KE-F-s-L}\\
-D\text{\ensuremath{\partial_{x}\left(d_{\omega}\partial_{x}F_{t}^{T}+j_{x,\omega}F_{t}^{L}\right)}}+ & \frac{d_{\omega}}{\tau_{s}}F_{t}^{T}=2\lambda N_{\omega}\delta(x)\left[P\left(n_{T}-F_{s}^{T}\right)-F_{t}^{T}\right]\label{KE-F-t-T}\\
-D\text{\ensuremath{\partial_{x}\left(\tilde{d}_{\omega}\partial_{x}F_{t}^{L}\right)-Dj_{x,\omega}\partial_{x}F_{t}^{T}}} & +\frac{1}{2}\Delta{\rm tr}\left\{ \tau_{1}(g^{R}+g^{A})\right\} F_{t}^{L}\nonumber \\
+\frac{d_{\omega}}{\tau_{s}}F_{t}^{L} & =2\lambda N_{\omega}\delta(x)\left[P\left(n_{L}-F_{s}^{L}\right)-F_{t}^{L}\right]\label{KE-F-t-L}
\end{align}
Here the coefficients are defined as follows
\begin{align}
d_{\omega}(x) & =\frac{1}{4}{\rm tr}\left\{ 1-g^{R}g^{A}\right\} \nonumber \\
N_{\omega}(x) & =\frac{1}{4}{\rm tr}\left\{ g^{R}-g^{A}\right\} \nonumber \\
\tilde{d}_{\omega}(x) & =\frac{1}{4}{\rm tr}\left\{ 1-\tau_{3}g^{R}\tau_{3}g^{A}\right\} \label{tilde-d}\\
j_{x,\omega}(x) & =\frac{1}{4}{\rm tr}\left\{ \tau_{3}\left(g^{R}\nabla_{x}g^{R}-g^{A}\nabla_{x}g^{A}\right)\right\} \equiv n_{\omega}(x)\partial_{x}\varphi\label{j-omega}\\
n_{\omega}(x) & =-\frac{i}{4}{\rm tr}\left\{ \tau_{3}g^{R}\tau_{3}g^{R}-\tau_{3}g^{A}\tau_{3}g^{A}\right\} \label{n-omega}
\end{align}
The phase gradient in these equation is determined from the selfconsistency
condition 
\begin{equation}
\int d\omega{\rm tr}\left\{ \tau_{2}g^{K}\right\} =\int d\omega{\rm tr}\left\{ \tau_{2}(g^{R}-g^{A})F_{s}^{L}-i\tau_{1}(g^{R}+g^{A})F_{s}^{T}\right\} =0\label{sefconsistency}
\end{equation}
which is equivalent to enforcing the charge conservation. Formally,
the condition \eqref{sefconsistency} ensures that after the energy
integration Eq. \eqref{KE-F-s-T} becomes the continuity equation.
This can be seen as follows. Let us trace Eqs.\eqref{g-R-equation}
and \eqref{g-A-equation} with $\tau_{3}$ and subtract them from
each other. The result is the following identity,
\[
D\partial_{x}j_{x,\omega}=\frac{i}{2}\Delta{\rm tr}\left\{ \tau_{2}(g^{R}-g^{A})\right\} 
\]
Using this identity I rewrite the selfconsistency condition in the
form
\begin{equation}
\frac{\Delta}{2}\int d\omega{\rm tr}\left\{ \tau_{1}(g^{R}+g^{A})F_{s}^{T}\right\} =-D\int d\omega F_{s}^{L}\partial_{x}j_{x.\omega}\label{selfconsistency-2}
\end{equation}
This equation guaranties that after the $\omega$--integration the
l.h.s. in Eq.\eqref{KE-F-s-T} becomes a total divergence, and can
be written as follows,
\[
\partial_{x}J(x)=I\delta(x)
\]
where $J(x)$ is the charge current in the wire and $I$ is the current
in the electrode,
\begin{align}
J(x) & =-\sigma_{D}\frac{1}{2}\int d\omega\tilde{d}_{\omega}(x)\partial_{x}F_{s}^{T}-\sigma_{D}\frac{1}{2}\int d\omega n_{\omega}(x)\partial_{x}\varphi { F_s^L}\label{J(x)}\\
I & =R_{b}^{-1}\frac{1}{2}\int d\omega N_{\omega}(x)\left[n_{T}-F_{s}^{T}-PF_{t}^{T}\right]\label{I-electrode}
\end{align}
Here $\sigma_{D}=DN_{F}$ is the Drude conductivity and $R_{b}=(2e^2\lambda N_{F})^{-1}$
is the contact resistance.

\subsection{Electric spin injection in normal metal}

Before considering the superconducting case, let us check the injection in  the normal state.
In this case,  $d_{\omega}=\tilde{d}_{\omega}=N_{\omega}=1,$
and $j_{x,\omega}=0$, and the  $T-$ and $L-$channels fully
decouple.   The equations for the distributions functions $F_{s}^{T}$
and $F_{t}^{T}$, Eqs. (\ref{KE-F-s-T}-\ref{KE-F-t-T}) simplify as, 
\begin{align}
-D\partial_{x}^{2}F_{s}^{T} & =2\lambda\delta(x)\left[n_{T}-F_{s}^{T}-PF_{t}^{T}\right]\label{KE-F-s-T-normal}\\
-D\partial_{x}^{2}F_{t}^{T}+\frac{1}{\tau_{s}}F_{t}^{T} & =2\lambda\delta(x)\left[P\left(n_{T}-F_{s}^{T}\right)-F_{t}^{T}\right]\label{KE-F-t-T-normal}
\end{align}
 One can integrate these equations over the energy
to get a closed system of equations for the charge and spin chemical
potentials,
\begin{align}
\mu & =\frac{1}{2}\int F_{s}^{T}(\omega)d\text{\ensuremath{\omega}}\label{mu-def-normal}\\
\eta & =\frac{1}{2}\int F_{t}^{T}(\omega)d\text{\ensuremath{\omega}}\label{eta-def-normal}
\end{align}
By multiplying Eqs. \eqref{KE-F-s-T-normal}-\eqref{KE-F-t-T-normal}
with $\frac{1}{2}e^{2}N_{F}$ and integrating over $\omega$, we get
\begin{align}
-\sigma_{D}\partial_{x}^{2}\mu(x) & =2e^{2}N_{F}\lambda\left[V-\mu(0)-P\eta(0)\right]\,\delta(x)\label{continuity-normal}\\
-\sigma_{D}\partial_{x}^{2}\eta(x)+\sigma_{D}\varkappa_{s}^{2}\eta(x) & =2e^{2}N_{F}\lambda\left[P\left(V-\mu(0)\right)-\eta(0)\right]\,\delta(x)\label{S-diffusion-normal-1}
\end{align}
Equation \eqref{continuity-normal} is the charge continuity equation.
 The coefficient in front of the delta--function in its
r.h.s. is the charge current flowing from the  F-electrode and $2e^{2}N_{F}\lambda=R_{b}^{-1}$ is  identified with the inverse boundary resistance of the contact. Thus, from Eq.\eqref{mu-def-normal}
we get the Ohm's law relating the current through the contact with
F-electrode to the voltage drop across the contact,
\begin{equation}
R_{b}I=V-\mu(0)-P\eta(0)\label{Ohm-law-normal}
\end{equation}
Finally, by expressing the source in Eg.\eqref{S-diffusion-normal-1}
in terms of the current $I$,  the spin diffusion equation reads:
\begin{equation}
-\sigma_{D}\partial_{x}^{2}\eta(x)+\sigma_{D}\varkappa_{s}^{2}\eta(x)=\left[PI-R_{b}^{-1}(1-P^{2})\eta(0)\right]\,\delta(x)\label{S-diffusion-normal-2}
\end{equation}
Eqs.\eqref{S-diffusion-normal-2}-\eqref{Ohm-law-normal} coincide in form  with  the equations used in the literature\cite{takahashi2003spin,sanz2020quantification,groen2023emergence}
This equations are valid in the tunneling contact limit.  In the case of transparent interfaces, in  the
source term in Eq.\eqref{S-diffusion-normal-2} one should replace $R_b$ by the  resistance of the ferromagnet $R_{s}^{F}=l_{s}^{F}/\sigma_{F}$ \cite{sanz2020quantification}.
In fact,  the inverse of the total resistance should
enter the second (back flow) term in the r.h.s. in Eq.\eqref{S-diffusion-normal-2}. 
{In our case,   $R_{b}\gg R_{s}^{F}$,  the large interface resistance suppresses the back flow, while
the source term, is  determined by the current, does not depend on
the transmission of the contact.}

\subsection{Electric spin injections in a superconductor: Linear regime}

The situation when L-- and T--channels decouple in general is the
linear response to the voltage bias $V.$ In this case,
\begin{align*}
n_{T} & =\frac{1}{2}\left[\tanh\left(\frac{\omega+V}{2T}\right)-\tanh\left(\frac{\omega+V}{2T}\right)\right]\approx\frac{V}{2T\cosh^{2}\left(\frac{\omega}{2T}\right)}\\
n_{L} & =\frac{1}{2}\left[\tanh\left(\frac{\omega+V}{2T}\right)+\tanh\left(\frac{\omega+V}{2T}\right)\right]=\tanh\left(\frac{\omega}{2T}\right)+O(V^{2})
\end{align*}
 Using these equalities  and inspecting Eqs.\eqref{KE-F-s-T}--\eqref{KE-F-t-L},
\eqref{j-omega} and \eqref{sefconsistency} one finds that
\begin{align*}
F_{s,t}^{T} & =O(V),\;j_{x,\omega}=O(V),\\
F_{s}^{L} & =\tanh\left(\frac{\omega}{2T}\right)+O(V^{2}),\;F_{t}^{L}=+O(V^{2})
\end{align*}
 Therefore the coupling can be  neglected and the kinetic equations
in the T--channel, read
\begin{align}
-D\text{\ensuremath{\partial_{x}\left(\tilde{d}_{\omega}\partial_{x}F_{s}^{T}\right)}}+\frac{\Delta}{2}{\rm tr}\left\{ \tau_{1}(g^{R}+g^{A})\right\} F_{s}^{T} & =2\lambda N_{\omega}\delta(x)\left[n_{T}-F_{s}^{T}-PF_{t}^{T}\right]\label{KE-F-s-T-linear}\\
-D\text{\ensuremath{\partial_{x}\left(d_{\omega}\partial_{x}F_{t}^{T}\right)+\frac{d_{\omega}}{\tau_{s}}F_{t}^{T}=}\,} & 2\lambda N_{\omega}\delta(x)\left[P\left(n_{T}-F_{s}^{T}\right)-F_{t}^{T}\right]\label{KE-F-t-T-linear}
\end{align}
where in the r.h.s. $n_{T}=\frac{V}{2T\cosh\left(\frac{\omega}{2T}\right)}$.
At $T\to0$ only $\omega=0$ contribution is relevant and this system
of equations possesses a complete analytic solution. 

\subsubsection{Solution of the injection problem}

In general, the solution to Eq.\eqref{KE-F-t-T-linear} can be constructed
in the same way as for the case of the spin--biased electrode. In
fact, Eq.\eqref{KE-F-t-T-linear} can be obtained from \eqref{KE-SC}
by the replacement $n_{T}\to P\left(n_{T}-F_{s}^{T}\right)$ in the
r.h.s. Therefore solution of Eq.\eqref{KE-F-t-T-linear} is immediately
obtained from \eqref{F(x)-approx},
\begin{equation}
F_{t}^{T}(x)=\frac{\lambda N_{\omega}(0)P\left(n_{T}-F_{s}^{T}(0)\right)}{Dd_{\omega}(0)\varkappa_{t}(\omega)+\lambda N_{\omega}(0)}e^{-\varkappa_{t}(\omega)|x|}\label{F-t-T-solution1}
\end{equation}
By inserting this result into the r.h.s. of \eqref{KE-F-s-T-linear}
we get a closed equation for the singlet distribution function $F_{s}^{T}$
which determines the charge imbalance,
\begin{align}
-D\text{\ensuremath{\partial_{x}\left(\tilde{d}_{\omega}\partial_{x}F_{s}^{T}\right)}}+\frac{\Delta}{2}{\rm tr}\left\{ \tau_{1}(g^{R}+g^{A})\right\} F_{s}^{T}\nonumber \\
=2\lambda N_{\omega}\frac{Dd_{\omega}\varkappa_{t}(\omega)+(1-P^{2})\lambda N_{\omega}}{Dd_{\omega}\varkappa_{t}(\omega)+\lambda N_{\omega}} & \left(n_{T}-F_{s}^{T}\right)\delta(x)\label{KE-s-linear-final}
\end{align}
{In this equation, in contrast to Eq.\eqref{KE-F-t-T-linear}, the coefficients in the l.h.s. are never small  , even
at $\lambda\to0$ ($\tilde d_\omega$ is of order unity at $\omega=0$). Therefore in  the tunneling contact limit, {\it i.e.}
large $R_{b}$ (small $\lambda$), the r.h.s. can be treated as a
perturbation. This means that $F_{s}^{T}(0)$ can be neglected compared to $n_{T}$ both in the r.h.s. of Eq.\eqref{KE-s-linear-final} and
in the r.h.s. of Eq.\eqref{F-t-T-solution1}. }In this regime we can relate the current
$I$ and the voltage $V$ in the injector simply by integrating the
r.h.s. of Eq.\eqref{KE-s-linear-final} over $\omega$,
\begin{equation}
R_{b}I=\frac{V}{2}\int d\text{\ensuremath{\omega}}N_{\omega}(0)\frac{Dd_{\omega}(0)\varkappa_{t}(\omega)+(1-P^{2})\lambda N_{\omega}(0)}{2T\cosh^{2}\left(\frac{\omega}{2T}\right)\left[Dd_{\omega}(0)\varkappa_{t}(\omega)+\lambda N_{\omega}(0)\right]}\label{I-V-injector}
\end{equation}
This relation can be written compactly as
\begin{equation}
V=R_{inj}I\label{V-I-injector}
\end{equation}
where the effective resistance of the injector is
\begin{equation}
R_{inj}^{-1}=\frac{R_{b}^{-1}}{2}\int d\text{\ensuremath{\omega}}N_{\omega}(0)\frac{Dd_{\omega}(0)\varkappa_{t}(\omega)+(1-P^{2})\lambda N_{\omega}(0)}{2T\cosh^{2}\left(\frac{\omega}{2T}\right)\left[Dd_{\omega}(0)\varkappa_{t}(\omega)+\lambda N_{\omega}(0)\right]}\label{R-inj}
\end{equation}
At the same level of accuracy the injected spectral spin reads,
\begin{equation}
S_{\omega}(x)=N_{\omega}(x)F_{t}^{T}(x)=PI\frac{R_{inj}\lambda N_{\omega}^{2}(0)e^{-\sqrt{\bar{\varkappa}_{\omega}^{2}+\varkappa_{s}^{2}}|x|}}{2T\cosh^{2}\left(\frac{\omega}{2T}\right)\left[Dd_{\omega}(0)\varkappa_{t}(\omega)+\lambda N_{\omega}(0)\right]}\label{S-F-injector}
\end{equation}

\subsubsection{Detection of the injected spin by magnetic detector}

Similarly to the spin--bias case in Sec.2.4, the voltage $V_{d}$
induced in the detector is calculated by setting to zero the detector
current 
\[
I_{d}\sim\int d\omega\left[N_{\omega}(x_{d})n_{T}(V_{d})-N_{\omega}(x_{d})F_{s}^{T}(x_{d})-P_{d}S_{\omega}(x_{d})\right]=0
\]
The important difference with Eq.\eqref{I-d} for the spin--biased
case is the presence of the charge imbalance componente, $F_{s}^{T}$,  at the location
of detector. This contribution is eliminated by subtracting the $V_{d}$
measured for parallel and antiparallel orientations of the injector
and detector polarizations, that is, $V_{d}^{\uparrow\uparrow}$ for
$P_{d}=P$, and $V_{d}^{\uparrow\downarrow}$ for $P_{d}=-P$. This is how non-local spin valves are operated\cite{jedema2001electrical,sanz2020quantification}
The
corresponding nonlocal resistance then reads,
\begin{equation}
R_{NL}=\frac{V_{d}^{\uparrow\uparrow}-V_{d}^{\uparrow\downarrow}}{I}=2P\frac{\int d\omega S_{\omega}(x_{d})}{I\int\frac{N_{\omega}(x_{d})d\omega}{2T\cosh^{2}\left(\frac{\omega}{2T}\right)}}\label{R-NL-def}
\end{equation}
By inserting here Eqs.\eqref{S-F-injector} and \eqref{R-inj} we
get explicitly,
\begin{equation}
R_{NL}=R_{b}P^{2}\frac{\int\frac{\lambda N_{\omega}^{2}(0)e^{-\sqrt{\bar{\varkappa}_{\omega}^{2}+\varkappa_{s}^{2}}x_{d}}d\omega}{2T\cosh^{2}\left(\frac{\omega}{2T}\right)\left[Dd_{\omega}(0)\varkappa_{t}(\omega)+\lambda N_{\omega}(0)\right]}}{\left[\int\frac{N_{\omega}(0)\left[Dd_{\omega}(0)\varkappa_{t}(\omega)+(1-P^{2})\lambda N_{\omega}(0)\right]d\omega}{4T\cosh^{2}\left(\frac{\omega}{2T}\right)\left[Dd_{\omega}(0)\varkappa_{t}(\omega)+\lambda N_{\omega}(0)\right]}\right]\left[\int\frac{N_{\omega}(0)e^{-\bar{\varkappa}_{\omega}x_{d}}d\omega}{4T\cosh^{2}\left(\frac{\omega}{2T}\right)}\right]}\label{R-NL-final}
\end{equation}
where $R_{b}=(2e^{2}\lambda N_{F})^{-1}$. This equation is used to plot $R_{NL}$ in the Fig. 2 (d) of teh main text.

In the limit $T\to0$ we
have $\frac{1}{2T\cosh^{2}(\omega/2T)}\to2\delta(\omega)$ and the
above expression simplifies dramatically,
\begin{equation}
R_{NL}=R_{b}P^{2}\frac{2\lambda e^{-\varkappa_{t}x_{d}}}{Dd_{0}(0)\varkappa_{t}+(1-P^{2})\lambda N_{0}(0)}\label{R-NL(0)}
\end{equation}
where $\varkappa_{t}=\sqrt{\varkappa_{0}^{2}+\varkappa_{s}^{2}}-\varkappa_{0}$.
From here the normal state result is obtained by setting $d_{0}=N_{0}=1$
and $\varkappa_{t}=\varkappa_{s}$, whereas in the superconductor
$d_{0}(0)=N_{0}^{2}(0)$ with $N_{0}(0)=\tilde{\lambda}=\lambda\xi_{0}/D=\frac{1}{2}\rho_{D}\xi_{0}/R_{b}$
(here $\rho_{D}=1/\sigma_D$ is the Drude resistivity). 

\section{Spin--to--charge conversion: Spin--galvanic effect}
\label{sec:s-2-c}
{Once we understood that spin can be injected even for voltages below the gap, we focus now on the spin-charge conversion via the spin-galvanic effect. For this we assume a sizable spin-orbit coupling in the superconductor. 
Moreover, because of its closer connection with experiments involving tunneling barriers, we assume here that $\eta/\Delta > \tilde{\lambda}$, such that the spectral functions of the superconductor, at zeroth order in the spin--galvanic parameter $\gamma$, are those given in Section~\ref{sec:Inj_Dynes}.}

We treat the problem perturbately in the spin-galvanic coeffcient $\gamma$\cite{kokkeler2025universal}.
The generated triplet GF now works as a  perturbation in the
singlet channel,
\[
-D\check{g}_{0}\left(\partial_{x}^{2}\check{g}_{1s}-i\partial_{x}^{2}\theta[\tau_{3},\check{g}_{0}]\right)+[-i\omega\tau_{3}+\Delta\tau_{1},\check{g}_{1s}]=-\partial_{x}\left(\gamma\check g_t\right)
\]
Importantly, the presence of the $\tau_{1}$ component in the Keldysh
source inevitably generates the $\tau_{2}$ component in the Keldysh
GF, which then breaks the continuity equation. To restore the charge
conservation we are forced to introduce a superconducting phase $\theta$
by $\check{g}\mapsto e^{-i\tau_{3}\theta}\check{g}e^{i\tau_{3}\theta}$,
determined from the condition
\begin{equation}
\label{eq:zeroimagDelta}
\int\frac{d\omega}{2\pi}{\rm tr}\left\{ \tau_{2}g^{K}(\omega)\right\} =0
\end{equation}
which ensures that there is no imaginary correction to $\Delta$,
and thus guaranties the charge conservation.
The correction, $g^R_1$, to the retarded GF is determined from the equation,
\[
-Dg_{0}^{R}\partial_{x}^{2}g_{1}^{R}+[-i\omega\tau_{3}+\Delta\tau_{1},g_{1}^{R}]=-iD\partial_{x}^{2}\theta g_{0}^{R}[\tau_{3},g_{0}^{R}]\; ,
\]
which in readily solved in the Fourier space,
\begin{equation}
g_{1}^{R}=\frac{iDq^{2}\theta_{q}}{Dq^{2}+2\Omega_{R}}[\tau_{3},g_{0}^{R}]\label{eq:g1R_sol}
\end{equation}
Similarly we find the advanced component.

The Keldysh component of the above equation reads,
\begin{align}
-D\left(g_{0}^{R}\partial_{x}^{2}g_{1s}^{K}+g_{0}^{K}\partial_{x}^{2}g_{1}^{A}\right)+[-i\omega\tau_{3}+\Delta\tau_{1},g_{1s}^{K}] & =-\partial_{x}\left(\gamma g_t^{K}\right)\label{eq:scc-Kcomp}\\
- & iD\partial_{x}^{2}\theta\left(g_{0}^{R}[\tau_{3},g_{0}^{K}]+g_{0}^{K}[\tau_{3},g_{0}^{A}]\right)\nonumber
\end{align}
 It is  convenient to write this equation using the standard representation of $g^{K}$ in terms of the distribution
function\cite{LarkinOvchinnikov1986},
\[
g^{K}=g^{R}F-Fg^{A}=g_{0}^{K}+(g_{1}^{R}-g_{1}^{A})h_{0}+(g_{0}^{R}\tau_{3}-\tau_{3}g_{0}^{A})h_{1T},
\]
where  the distribution function has the form, $F=h_{0}+\tau_{3}h_{1T}$
with $h_{0}=\tanh\frac{\omega}{2T}$. Thus, from Eq. (\ref{eq:scc-Kcomp}) we find the the equation for the non-equlibrium part
of the distribution function,
\begin{equation}
-D\left(\tau_{3}-g_{0}^{R}\tau_{3}g_{0}^{A}\right)\partial_{x}^{2}h_{1T}+[-i\omega\tau_{3}+\Delta\tau_{1},g_{0}^{R}\tau_{3}-\tau_{3}g_{0}^{A}]h_{1T}=-\partial_{x}\left(\gamma g_t^{K}\right)\; ,
\end{equation}
which can be  further simplified,
\begin{equation}
\left(\tau_{3}-g_{0}^{R}\tau_{3}g_{0}^{A}\right)\left\{ -D\partial_{x}^{2}h_{1T}+(\Omega_{R}+\Omega_{A})h_{1T}\right\} =-\partial_{x}\left(\gamma g_t^{K}\right)
\end{equation}
Notice that the matrix structure on either side of this equation is
identical, which confirms the ansatz for the distribution function.
The explicit solution takes the form,
\begin{equation}
    \label{eq:h1T_sol}h_{1T}=\gamma\frac{-iq2\lambda_{q}\tau_{s}n_{T}(\omega)}{\left[Dq^{2}+\Omega_{R}+\Omega_{A}\right]\left[1+\tau_{s}(\Omega_{R}+\text{\ensuremath{\Omega_{A}}})+l_{s}^{2}q^{2}\right]}
\end{equation}

\subsection{Charge imbalance and the voltage drop}

The induced distribution function $h_{1T}$ determines the charge
imbalance induced in the superconductor,
\[
Q_{q}^{*}=\frac{\pi\nu}{4}\int\frac{d\omega}{2\pi}{\rm tr}\left\{ g^{K}(\omega)\right\} =\frac{\nu}{2}\text{\ensuremath{\int d\omega N_{0}(\omega)h_{1T}=\gamma\frac{\nu\lambda_{q}\tau_{s}}{iDq\left(1+l_{s}^{2}q^{2}\right)}}}\int d\omega N_{0}(\omega)n_{T}(\omega)
\]
and the potential drop across the injector,
\begin{equation}
e\Delta\varphi=i\lim_{q\to0}qQ_{q}^{*}=\gamma\frac{\nu\lambda\tau_{s}}{D}\int d\omega N_{0}(\omega)n_{T}(\omega)
\end{equation}
 Notice that at $T\to0$ and $V_{s}<\Delta$ no
charge imbalance and no voltage drop {(larger than $\eta$)} is induced from the injected
spin. In this situation only a supercurrent and or a phase drop are
expected.

\subsection{Phase gradient and the phase drop}\label{sub-section:phase-drop}

The phase gradient is determined by the condition, Eq(\ref{eq:zeroimagDelta}),
\[
\int\frac{d\omega}{2\pi}{\rm tr}\left\{ \tau_{2}g_{1s}^{K}\right\} =\int\frac{d\omega}{2\pi}{\rm tr}\left\{ \tau_{2}(g_{1}^{R}-g_{1}^{A})h_{0}+\tau_{2}(g_{0}^{R}\tau_{3}-\tau_{3}g_{0}^{A})h_{1T}\right\} =0
\]
or, equivalently, by the requirement that the continuity equation
is fulfilled,
\[
\partial_{x}j(x)=0
\]
where the current is defined as follows,
\begin{align*}
j & =-\frac{\pi N_F}{2}D\int\frac{d\omega}{2\pi}{\rm tr}\left\{ \tau_{3}\left[\check{g}\partial_{x}\check{g}-i\partial_{x}\theta\check{g}[\tau_{3},\check{g}]\right]^{K}\right\} +\frac{\pi N_F }{2}\int\frac{d\omega}{2\pi}{\rm tr}\left\{ \tau_{3}\gamma g_{t}^{K}\right\} \\
=- & \frac{N_F }{4}D\partial_{x}\int d\omega{\rm tr}\left\{ \left(1-\tau_{3}g_{0}^{R}\tau_{3}g_{0}^{A}\right)h_{1T}+h_{0}\tau_{3}\left[g_{0}^{R}\left(g_{1}^{R}-i\theta[\tau_{3},g_{0}^{R}]\right)-g_{0}^{A}\left(g_{1}^{A}-i\theta[\tau_{3},g_{0}^{A}]\right)\right]\right\} \\
 & +\frac{N_F }{4}\int d\omega{\rm tr}\left\{ \tau_{3}\gamma g_t^{K}\right\} 
\end{align*}
With the solutions for $g_{1}^{R,A}$ and $h_{1T}$ from Eqs. (\ref{eq:g1R_sol},\ref{eq:h1T_sol}), the Fourier component
of the current is given by
\begin{align*}
j_{q} & =\gamma\frac{N_F }{2}\int d\omega\frac{\Omega_{R}+\Omega_{A}}{Dq^{2}+\Omega_{R}+\Omega_{A}}s_{q}(\omega)\\
+q\theta_{q}\frac{\sigma_{0}}{4}\int d\omega h_{0}(\omega) & \left[\frac{\Omega_{R}}{Dq^{2}+\Omega_{R}}{\rm tr}\left\{ \tau_{3}g_{0}^{R}[\tau_{3},g_{0}^{R}]\right\} -\frac{\Omega_{A}}{Dq^{2}+\Omega_{A}}{\rm tr}\left\{ \tau_{3}g_{0}^{A}[\tau_{3},g_{0}^{A}]\right\} \right]\;.
\end{align*}
which after evaluation of the traces result in, 
\begin{equation}
\label{eq:jq_sol}
j_{q}=\gamma\frac{N_F }{2}\int d\omega\frac{\Omega_{R}+\Omega_{A}}{Dq^{2}+\Omega_{R}+\Omega_{A}}s_{q}(\omega)-iq K_{s}(q)\theta_{q}
\end{equation}
where we have  introduced a $q-$dependent superfluid weight,
\begin{align*}
K_{s}(q) & =-i\sigma_{0}\int d\omega\left\{ \frac{\Delta^{2}}{(Dq^{2}+\Omega_{R})\Omega_{R}}-\frac{\Delta^{2}}{(Dq^{2}+\Omega_{A})\Omega_{A}}\right\} \text{\ensuremath{\tanh\frac{\omega}{2T}}}\\
 & =4\pi\sigma_{0}T\sum_{\omega_{n}}\frac{\Delta^{2}}{\left(Dq^{2}+\sqrt{\Delta^{2}+\omega_{n}^{2}}\right)\sqrt{\Delta^{2}+\omega_{n}^{2}}}
\end{align*}

The phase $\theta$ is determined by the condition $\partial_{x}j(x)=0$,
which requires $j(x)=const$.  The constant depends on the geometry
and boundary conditions. { There are two relevant situations, sketeched in Fig. 1(c-d) of the main text and which we analyze next.}

I. \textbf{\emph{Wire with open boundaries }(Fig. 1c in main text) :} In this case the current
should be identically zero $j(x)=0$ which determines the phase distribution,
\[
iq K_{s}(q)\theta_{q}=\gamma N_F \int_{0}^{\infty}d\omega\frac{\Omega_{R}+\Omega_{A}}{Dq^{2}+\Omega_{R}+\Omega_{A}}s_{q}(\omega)
\]
For the induced anomalous phase drop across the injector we get depending
of the wire length $L$,
\[
\delta\theta=i\lim_{q\to0}q\theta_{q}=\begin{cases}
\frac{\gamma}{K_{s}}N_F \int_{0}^{\Delta}d\omega s_{q=0}(\omega)=\frac{\gamma}{K_{s}(0)}S_{\Delta}\;, & L<\Lambda_{Q^{*}}\\
\frac{\gamma}{K_{s}}N_F \int_{0}^{\infty}d\omega s_{q=0}(\omega)=\frac{\gamma}{K_{s}(0)}S\;, & L>\Lambda_{Q^{*}}
\end{cases}
\]
where $\Lambda_{Q^{*}}$ is the charge imbalance length\cite{beckmann2016spin}.
Formally, the first case corresponds to taking the limit $\eta\to0$
before the limit $q\to0$, and the in the second case one takes $q\to0$
first while keeping $\eta$ finite.

II. \textbf{\emph{Closed loop geometry}(Fig. 1d in main text):} In this case the current
$j(x)=I$ induced in the loop is given by the $q=0$ component of
the  current density Eq. (\ref{eq:jq_sol}) under the condition of regularity and periodicity
of the phase, $I\cdot L=j_{q=0},$ which implies, 
\begin{equation}
I=\frac{\gamma}{L}N_F \int_{0}^{\infty}d\omega s_{ q=0}(\omega)=\frac{\gamma}{L}S\label{eq:I_sol}
\end{equation}
The total current is thus given by the total injected spin divided
by the loop length. This is the same as in the normal state, but the
current is larger, because the injected spin is larger in the superconducting case , and  part
of it is superfluid. The distribution of the supercurrent along the
loop is readily found as,
\begin{equation}
j_{s}(q)\equiv iq\varkappa_{s}(q)\theta_{q}=\gamma N_F \int_{0}^{\infty}d\omega\frac{Dq^{2}}{Dq^{2}+\Omega_{R}+\Omega_{A}}s_{q}(\omega)
\end{equation}
{ We have ignored the magnetic field induced by the current, assuming that the current is small compared to the critical current.}





\end{document}